\documentclass[11pt, a4paper, oneside]{article}
\usepackage[body={15.5cm, 24.0cm},left=2cm,right=2cm]{geometry}
\usepackage{natbib}
\usepackage{amsmath, amssymb}
\usepackage{nicefrac}
\usepackage{utopia}
\usepackage{graphicx}
\usepackage[usenames]{color}
\usepackage{setspace}
\usepackage{pifont}                  
\usepackage{hyperref}
\usepackage{natbib}
\usepackage{color,soul}
\usepackage{mathtools}
\usepackage{dsfont} 

\usepackage{epstopdf}
\usepackage{schemata}
\usepackage{enumitem}
\usepackage{algorithm}
\usepackage{algpseudocode}
\usepackage{booktabs}

\usepackage{authblk}

\usepackage{tikz,xcolor,hyperref}
\definecolor{lime}{HTML}{A6CE39}
\DeclareRobustCommand{\orcidicon}{%
	\begin{tikzpicture}
	\draw[lime, fill=lime] (0,0) 
	circle [radius=0.16] 
	node[white] {{\fontfamily{qag}\selectfont \tiny ID}};
	\draw[white, fill=white] (-0.0625,0.095) 
	circle [radius=0.007];
	\end{tikzpicture}
	\hspace{-2mm}
}

\foreach \x in {A, ..., Z}{%
	\expandafter\xdef\csname orcid\x\endcsname{\noexpand\href{https://orcid.org/\csname orcidauthor\x\endcsname}{\noexpand\orcidicon}}}

\numberwithin{equation}{section}

\newcommand{\field}[1]{\mathbb{#1}}
\newcommand{\real}{\ensuremath{{\field{R}}}}

\newcommand{\sumab}[2]{\ensuremath{\sum\limits_{#1}^{#2}}}

\newcommand{\argmin}[1]{\ensuremath{\displaystyle {\arg\min_{#1}}}}

\newcommand\minover[1]{\underset{#1}{\mathrm{min} \text{ }}}

\newcommand{\HsSP}{$H_S^{sp}$} 
\makeatletter
\newcommand*\bigcdot{\mathpalette\bigcdot@{.5}}
\newcommand*\bigcdot@[2]{\mathbin{\vcenter{\hbox{\scalebox{#2}{$\m@th#1\bullet$}}}}}
\newcommand{\one}{\mathds{1}}

\newcommand{\keywords}[1]{{\scriptsize \noindent \textbf{KEY WORDS AND PHRASES:}} {#1}\\}

\newcommand{\un}[1]{\boldsymbol{#1}}

\title{Modelling non-stationary extremes of storm severity:\\ a tale of two approaches}
\author[a]{Evandro Konzen}
\author[a]{Cl\'{a}udia Neves}
\author[b,c]{Philip Jonathan}
\affil[a]{{\small Department of Mathematics and Statistics, University of Reading}}
\affil[b]{{\small Shell Research Ltd.}}
\affil[c]{{\small Department of Mathematics and Statistics, Lancaster University}}
\date{}

\begin{document}

\maketitle

\strut

\abstract{Models for extreme values accommodating non-stationarity have been amply studied and evaluated from a parametric perspective. Whilst these models are flexible, in the sense that many parametrizations can be explored, they assume an asymptotic distribution as the proper fit to observations from the tail. This paper provides a holistic approach to the modelling of non-stationary extreme events by iterating between parametric and semi-parametric approaches, thus providing an automatic procedure to estimate a moving threshold with respect to a periodic covariate in circular data. By exploiting advantages and mitigating pitfalls of each approach, a unified framework is provided as the backbone for estimating extreme quantiles, including that of the $T$-year level and finite right endpoint, which seeks to optimize  bias-variance trade-off. To this end, two tuning parameters related to the spread of peaks over threshold are introduced. We provide guidance for applying the methodology to the directional modelling of hindcast storm peak significant wave heights recorded in the North Sea. Although the theoretical underpinning for adaptation of well-known estimators in statistics of extremes to circular data is given in some detail, the derivation of their asymptotic properties lays  beyond the scope of this paper. A bootstrap technique is implemented for obtaining direction-driven confidence bounds in such a way as to account for the relevant boundary restrictions with minimal sensitivity to initial point. This provides a template for other applications where the analysis of directional extremes is of importance.}

\vspace{0.2cm}



\section{Introduction and Motivation} \label{Sct:Int}

Statistical inference for extreme values has been a dynamic and rapidly developing field over the last decade or so, and offers considerable scope for practical application in science and engineering. Curiously, in the midst of this active development, two seemingly divergent camps of statistical thought have emerged, proposing different approaches to extreme value modelling, yielding inferences which are not always obviously in agreement. The work presented in this paper seeks to identify potential points of contact between these so-called parametric and semi-parametric frameworks for extreme value inference, to encourage better common understanding and convergence of at least some practices, in particular for tackling non-stationary extremes.

Non-stationarity is commonplace in environmental extremes; physical processes generate extreme values which typically vary systematically with covariates, including space and time. For the peaks-over-threshold (POT) method, where only data exceeding a threshold are used for analysis \citep{Balkema1974,Pickands1975}, various models have been proposed to capture non-stationarity, including those of \citet{davison1990models} and \citet{leadbetter91}. In the parametric framework, non-stationarity can be incorporated within the appropriate distribution function for threshold exceedances by allowing the distribution's (shape and scale) parameters to vary with covariates \citep[see e.g.][and references therein]{coles2001introduction,chavez2005gamExtremes}. Important assumptions underpinning this approach are that the data generating process is locally stationary and that observations from the data generating process can be considered approximately independent given covariates. 

The need for non-stationary extreme value threshold is well-recognised for environmental applications. \citet{RbsTwn97} base their approach on a non-constant threshold to characterise the evolution of extreme sea currents. \citet{northrop2011threshold} propose a covariate-dependent threshold estimated using quantile regression, and  \citet{northrop2016thre} propose a cross-validation procedure for threshold selection. We introduce a method for selection of a non-stationary threshold amenable to both parametric and semi-parametric approaches to inference. The basic  assumption is that the shape parameter $\xi$ (the key parameter for quantifying  tail-heaviness) does not depend on the covariate. This assumption is the starting point for conciliation between parametric and semi-parametric approaches.

Applications of non-stationary extreme value analysis are more numerous within the remit of parametric inference than in the semi-parametric setting. For example, in an ocean engineering context, \citet{forristall2004use} performs extreme value analysis of significant wave height for directional octants. This approach (i) accommodates directional non-stationarity and (ii) allows extreme quantiles for specific directional sectors to be estimated. Choice of number and widths of directional sectors is an open problem (see e.g. \citealt{RssEA17b, folgueras2019selection}). These choices constitute a difficult problem as the environmental extremes usually change smoothly with respect to direction, motivating use of various basis representations for parameters of the conditional distribution of threshold exceedances, as a function of covariate (see e.g.  \citealt{jones2016statistics, ZnnEA19a}). 

The goal in this paper is to combine parametric and semi-parametric modelling approaches to obtain a new method for inference on circular extreme data. Firstly, we augment the scope of  the semi-parametric approach so that semi-parametric inference for quantities of interest in ocean engineering is possible and meaningful. This will be achieved via an adaptive method for threshold selection. We then present a comparative survey showing how the parametric approach can be complemented with semi-parametric methodology aiming at improved inference for directional extremes. In particular, we propose a unified procedure for inference that borrows insight from both frameworks and harmonising between them in terms of (i) model-fit, and (ii) estimated extreme value indices such as the shape parameter or extreme value index, $T$-year value or an extreme quantile including the right endpoint of the support of the underlying distribution. Finally, the main application for illustration of our approach involves directional extreme value analysis of hindcast storm peak significant wave height (henceforth referred to as \HsSP{}) recorded at a northern North Sea location offshore Norway. We hope to demonstrate that the proposed approach to non-stationary extreme value analysis may be of practical benefit to practitioners in coastal and ocean engineering and environmental sciences.

The remainder of the paper is organized as follows. The motivating application to directional modelling of \HsSP{} is introduced in Section \ref{Sct:Dat}. Section \ref{Sct:ThrStt} provides key definitions and theoretical results underpinning the combined methodology for threshold selection in the presence of non-stationarity that will be developed in Section \ref{Sct:ThrEst}. Sections \ref{Sct:SplML} and \ref{Sct:ExtQnt} detail the adapted estimators in this paper. Our main application of the parametric and semi-parametric combined methodology is presented in Section \ref{Sec:App}. Finally, Section \ref{Sec:Summary} lists the main contributions of the work.

\section{Motivating application}\label{Sct:Dat}

\begin{figure}
	\centering
	\includegraphics[scale=0.2]{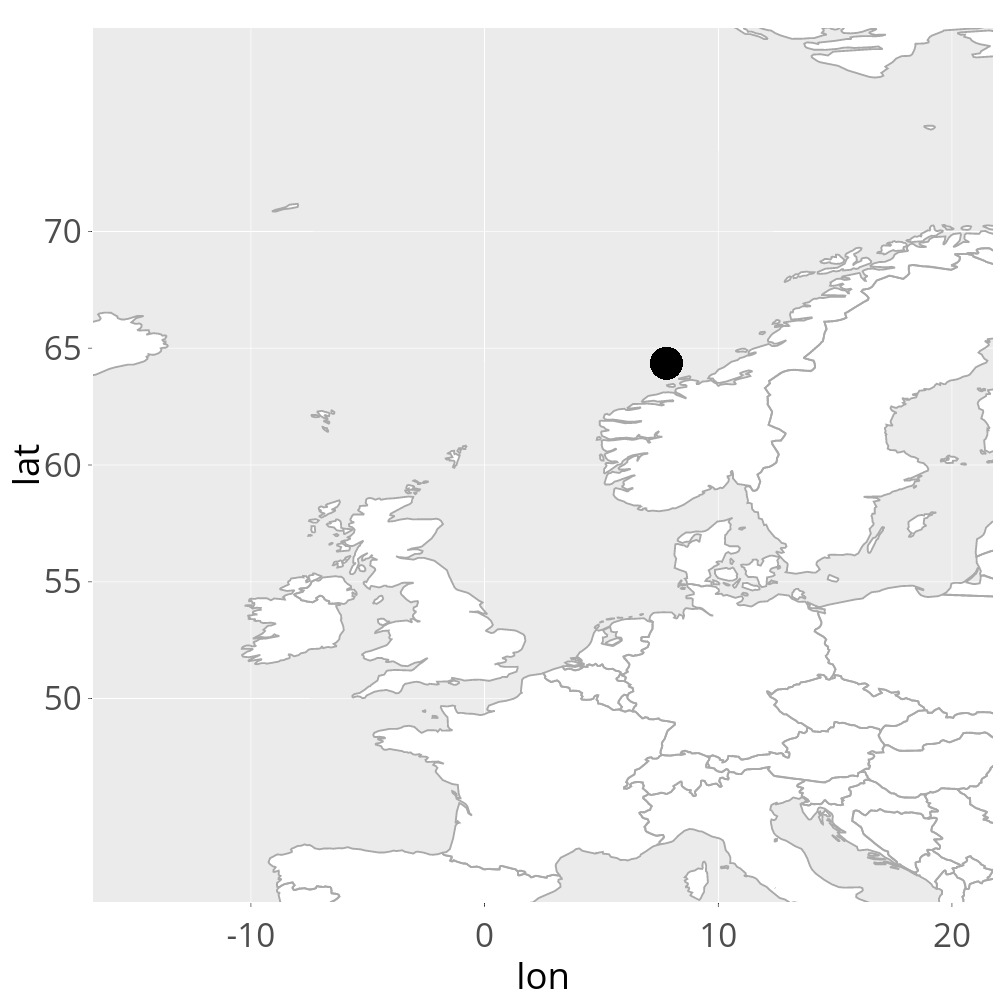}\\
	\includegraphics[scale=0.5]{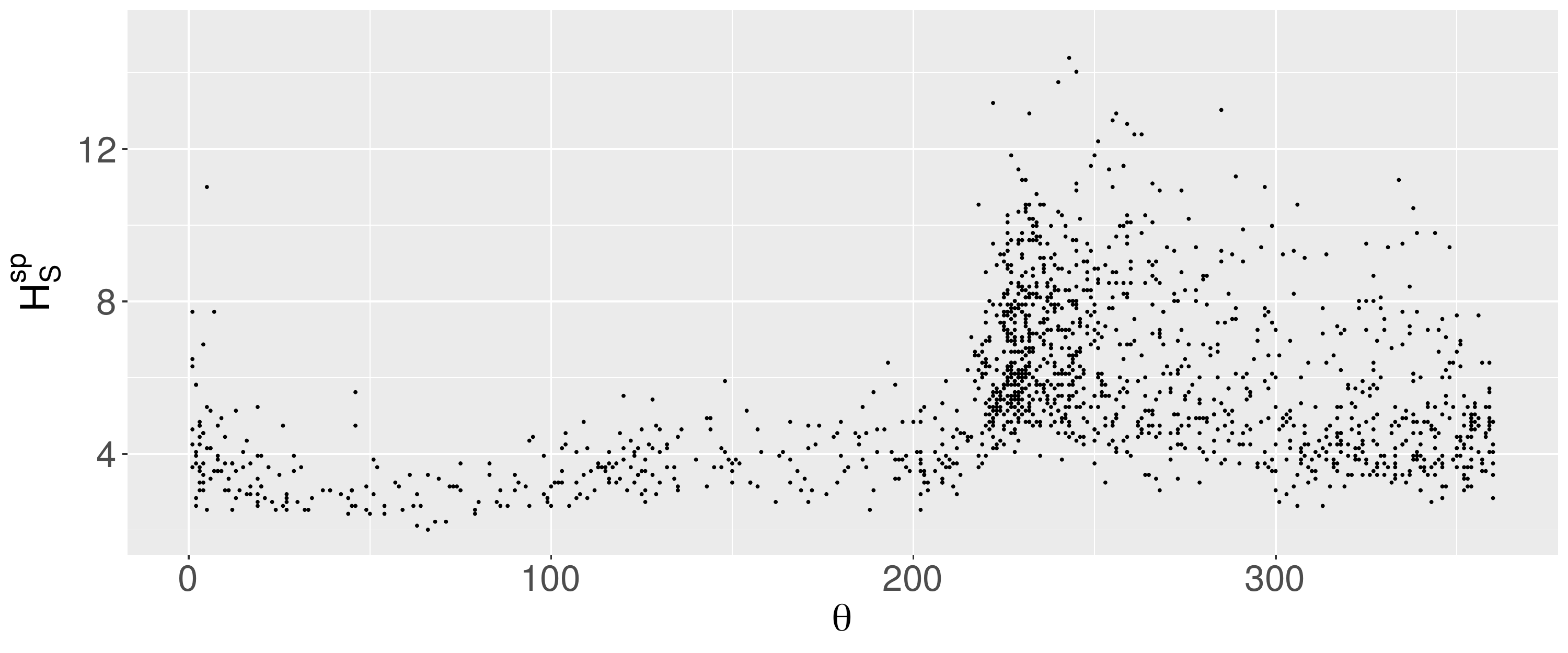}
	\caption{~Map showing the relevant location at North Sea \emph{(top)}; scatter-plot displaying the $H_S^{sp}$ data against 360 directions from which wave propagate, measured clockwise from north, expressed in the angular component $\theta= 0,1, \ldots, 359^\circ$\emph{(bottom)}.}
	\label{Fig:HsSPdata}
\end{figure}

The sample of data for the motivating application is described by \cite{RndEA15a}. The data corresponds to observations of storm severity and storm direction in the northern North Sea. Significant wave height (H$_{\mbox{s}}$) measures the roughness of the ocean surface, and can be defined as four times the standard deviation of the ocean surface elevation at a spatial location for a specified period of observation. The application sample is taken from the WAM hindcast of \cite{RstEA11}, which provides time-series of significant wave height, (dominant) wave direction and season (defined as day of the year, for a standardised year consisting of 360 days) for three hour sea-states for the period September 1957 to December 2012 at a northern North Sea location in the vicinity of the black disk in the upper panel of Figure \ref{Fig:HsSPdata}. A hindcast is a physical model of the ocean environment, incorporating pressure field, wind field and wind-wave generation models in particular; the hindcast model is calibrated to observations of the environment from instrumented offshore facilities, moored buoys and satellite altimeters in the neighborhood of the location for a period of time, typically decades. Extreme seas in the North Sea are dominated by winter storms originating in the Atlantic Ocean and propagating eastwards across the northern part of the North Sea. Due to their proximity to the storms, sea states at northern North Sea locations are usually more intense than in the southern North Sea. Occasionally, the storms travel south-eastward and intrude into the southern North Sea producing large sea states. Directions of propagation of extreme seas vary considerably with location, depending on land shadows of the British Isles, Scandinavia, and the coast of mainland Europe, and fetches associated with the Atlantic Ocean, Norwegian Sea, and the North Sea itself. In the northern North Sea the main fetches are the Norwegian Sea to the North, the Atlantic Ocean to the west, and the North Sea to the south. Extreme sea states from the directions of Scandinavia to the east and the British Isles to the south-west are not possible. The shielding by these land masses is more effective for southern North Sea locations, resulting in a similar directional distribution but reduced wave heights by comparison with northern North Sea locations.

At the location of interest, observations of storm peak significant wave height (\HsSP) are isolated from the hindcast time-series using the procedure described in \cite{EwnJnt08} as follows. Briefly, contiguous intervals of significant wave height above a low peak-picking threshold are identified, each interval assumed to correspond to a storm event. The peak-picking threshold corresponds to a directional-seasonal quantile of H$_{\mbox{s}}$ with specified non-exceedance probability, estimated using quantile regression. The maximum of significant wave height during the storm interval is taken as the storm peak significant wave height \HsSP{}. The values of directional and seasonal covariates at the time of storm peak significant wave height are referred to as storm peak values of those variables. The resulting storm peak sample consists of 2941 values. With direction \emph{from which} a storm travels expressed in degrees clockwise with respect to north, the lower panel of Figure \ref{Fig:HsSPdata} shows a polar plot of observations of \HsSP{} (in metres) versus direction.

The land shadow of Norway (approximately the directional interval $(45^\circ,210^\circ)$) has a considerable effect on the rate and size of occurrences with direction. In particular, there is a dramatic increase in both rate and size of occurrences with increasing direction at around $210^\circ$, corresponding to Atlantic storm events from the south-west able to pass the Norwegian headland. We therefore should expect considerable directional variability in model parameter estimates for the sample. In contrast, the magnitude of the rate of change of both rate and size of occurrences with respect to season (not shown, but see \citealt{RndEA15a}) is lower. Winter storms (approximately from October to March) are more intense and frequent. Only winter storms with storm peak events occurring in October to March including have been considered further in this work which corresponds to 1521 data points.

\section{Theoretical motivation for the stationary case}\label{Sct:ThrStt}
Here we summarise basic theoretical results and assumptions underpinning extreme value modelling for the stationary case. Further details are given in the Appendix. Extension to non-stationary will be present in Section~\ref{Sct:SplML}.

Suppose the available sample consists of realizations of a sequence of independent and identically distributed (i.i.d)  random variables $X_1, X_2, \dots, X_n$. The sequence can also be weakly dependent. We assume that all the random variables follow the same (unknown) distribution function, and for brevity use the symbol $X$ to refer to any of the random variables when there is no need to be more specific. The common distribution function is $F(x) = P(X \leq x)$, for every $x \in \mathbb{R}$. We denote by $x^F$ the right endpoint of the support of $F$, namely the ultimate value which bounds all possible observations from above,
\begin{equation}\label{EP}
x^F= \sup\{x:\, F(x)<1\}.
\end{equation}
We note that $x^F$ may be less than or equal to infinity.

The extreme value (or extreme types) theorem (\citealt{FisherTippett:28,Gnedenko:43,deHaan:70}) establishes that the limit distribution of linearly normalised partial maxima $M_n= \max(X_1,  \dots, X_n)$, with real constants $a_n>0$ and $b_n\in \mathbb{R}$, must be one of only possible three extreme value distributions: Fr\'echet, Gumbel or Weibull. These three types can be nested in the one-parameter Generalized Extreme Value (GEV) distribution. Specifically, if there exist $a_n>0$ and $b_n\in \mathbb{R}$ such that,
\begin{equation}\label{DOA}
\lim_{n\rightarrow \infty}P\big\{a_n^{-1}(M_n-b_n)\leq x\big\}= \lim_{n\rightarrow \infty} F^n(a_n x+b_n) = G(x),
\end{equation}
for every continuity point of (non-degenerate) $G$, then $G$ must be a GEV distribution with distribution function  given by:
\begin{equation}\label{DistNormMax}
G_{\xi}(x)= \exp \bigl( -(1+\xi\, x)^{-1/\xi} \bigr),
\end{equation}
for all $x$ such that $1+\xi x >0$. We then say that $F$ belongs to the max-domain of attraction of the GEV, for some $\xi\in \real$, and write $F\in \mathcal{D}(G_\xi)$.  Parameter $\xi$ is conventionally referred to as the shape parameter in parametric literature, and the extreme value index (EVI) in semi-parametric literature of extremes. The Fr\'{e}chet max-domain of attraction, corresponding to $\xi>0$, contains distributions exhibiting polynomial decay, such as the Pareto, Cauchy, Student's $t$ and Fr\'{e}chet itself. These distributions have infinite right endpoint $x^F$. All distribution functions belonging to $\mathcal{D}(G_\xi)$ with $\xi<0$, referred to as the Weibull max-domain of attraction, are short-tailed with finite $x^F$; examples include the uniform and beta distribution. For $\xi=0$, a continuity  argument gives  $G_{\xi=0}(x)=\exp\bigl(-e^{-x}\bigr)$, $x\in \mathbb{R}$. This  gives rise to the Gumbel max-domain of attraction ($\mathcal{D}(G_\xi)$ with $\xi=0$, or $\mathcal{D}(G_0)$), a domain of particular interest in many applied fields, due to both simplicity of inference and the great variety of distributions possessing an exponential tail (with either $x^F < \infty$ or $x^F = \infty$). The normal, gamma and log-normal distributions are only a few members of $\mathcal{D}(G_0)$.

Extreme value inference using block maxima has been practised for many years, from the time of \cite{Gumbel58} in application to hydrology. However, in this paper, inference is based on analysis of threshold exceedances (or POT), motivated by characterization of the max-domains of attraction above in terms of exceedances of high threshold. \citet{Balkema1974} and \citet{Pickands1975} established that the max-domain of attraction condition \eqref{DOA} is equivalent to the assertion that the (conditional) distribution of $X$ given $X>u$, with $u$ near the right endpoint $x^F$, converges to the Generalized Pareto distribution (GPD) with distribution function $1+\log G_\xi$. Analysis of threshold exceedances (or POT) is potentially statistically more efficient than analysis of block maxima: in the former, all large values of threshold exceedance in the sample are admitted, including multiple occurrences of large values belonging to same block, which might be excluded in a block maximum analysis. 

The parametric and semi-parametric approaches to extreme value analysis considered in this paper have strong conceptual similarities. Both approaches involve estimation of three quantities. In the parametric setting, we estimate GPD shape and scale parameters, the latter being threshold-dependent; for estimation of T-year levels, we also need to estimate the rate of threshold exceedance. In the semi-parametric analysis, we focus on estimation of a single parameter, the so-called extreme value index (i.e. the semi-parametric equivalent of the GPD  shape parameter); associated scale and location normalising functions (akin to $a>0$ and $b$ in \eqref{DOAnonstat}  on the real line) are estimated separately. 

Sensitivity to the extreme value threshold choice is a common critical feature of the approaches. Confirming the relative stability of estimated extreme value index (or shape parameter) and high quantiles (such as the $T$-year level), with near-zero exceedance probability $p$ with respect to threshold is a key diagnostic test for the analysis. This will be explained at length in Section \ref{Sct:ExtQnt}.

\section{Non-stationary threshold selection}\label{Sct:ThrEst}
We extend the univariate setting of Section~\ref{Sct:ThrStt} to the non-identically distributed case as follows. We assume that the covariate domain $\mathcal{S}$ is partitioned into $m$ intervals, with centroids $\theta_j$, and write $\Theta$ for the set $\{\theta_j\}_{j=1}^m$ of centroids. Specifically for directional analysis considered here, $\mathcal{S}=[0^\circ,360^\circ)$, and $\theta_j=j-1$, $j=1,...,360$. Suppose, for each $j$, that $X_1(\theta_j),\,  \dots,\, X_{n}(\theta_j)$  consists of a sample of $n$ (unknown) independent identically-distributed (positive) random variables.  We assume the extreme value theorem holds at each $\theta_j$, for sufficiently large $n$. The extreme value condition \eqref{DOA} shows that the shape parameter (or extreme value index) $\xi(\theta_j)$ governs the tail behaviour of the underlying distribution {$F_{\theta_j}(x)=P\{X(\theta_j) \leq x\}$, $x\in \real$}. In the non-stationary case, our objective is estimation of {$\xi(\theta_j)$} (and associated parameters discussed in Section~\ref{Sct:ThrStt}) from a sample of threshold exceedances. We achieve this parametrically using maximum likelihood estimation for the GPD \citep{coles2001introduction, davison1990models}, and semi-parametrically using the moment estimator \citep{dekkers1989moment}. Enhanced versions of these estimators, designed to accommodate non-stationarity, will be introduced later on in this section.

Extreme value statistics characterises the largest values in the sample and therefore choosing an appropriate threshold above which inference will take place is as essential as choosing between inference approaches.  As noted in Section~\ref{Sct:Int}, numerous authors have considered the estimation of non-stationary thresholds for extreme value analysis based on POT-GPD. Indeed, it appears sensible to consider a covariate-dependent threshold for inference using data showing strong covariate dependence. Motivated by the desire to admit the same proportion of the original sample for subsequent extreme value analysis given covariate, threshold estimation using a covariate-conditional quantile regression would seem an obvious choice \citep{northrop2011threshold, northrop2016thre}. Using this approach, a constant threshold exceedance probability of $\tau$  given covariate might be sought. However, our studies have demonstrated that this approach might not strike the right bias-variance balance for efficient parameter estimation.

Section~\ref{Sct:ThrEst:Hrs} below introduces our heuristic criterion devised for threshold selection, which relies on local estimates for $\xi$ on the covariate domain. These can be inferred using semi-parametric (Section~\ref{Sct:ThrEst:SP}) or maximum likelihood estimation (Section~\ref{Sct:ThrEst:ML}). Section~\ref{Sct:ThrEst:Alg} provides a simple algorithm for the threshold selection procedure.

\subsection{Heuristic criterion for threshold selection} \label{Sct:ThrEst:Hrs}
Let $\mathcal{N}(\theta_j, h)$ be the directional neighborhood with  center at $\theta_j$ and with fixed radius $h>0$ defined by:
\begin{equation}\label{NeigbDirect}
	\mathcal{N}(\theta_j, h) = \{{\theta^* \in \mathcal{S}}:\,  0\leq d(\theta^*, \theta_j) \leq h \},
\end{equation}
for every $\theta_j \in \Theta$, equipped with the wrapped-Euclidean distance on $\mathcal{S}$:
\begin{equation}\label{circDist}
	{d(\theta^*, \theta_j) := \min \big\{ |\theta^* - \theta_j|, 360 - |\theta^* - \theta_j|\big\}.}
\end{equation}
We propose a heuristic threshold selection procedure that hinges on the propensity for extreme values to concentrate in the neighborhood $\mathcal{N}(\theta_j, h)$ of the centroid $\theta_j \in \Theta$. For each $\theta_{j}= j-1$ defined above, a threshold $u(\theta_j)$ is  set automatically by drawing on the realisations of $X_i (\theta_{j'})$, $i=1, \ldots, n$, $\theta_{j'} \in \Theta \subset \mathcal{S}$, within lag $h$ of centroid $\theta_j$ resulting in a tally of:
\begin{equation*}
N(\theta_j)= \sumab{i=1}{n}\sumab{j'=1}{360} \one_{\{\theta_{j'} \in \mathcal{N}(\theta_j, h)\}}\bigl( X_i(\theta_{j'}) \bigr)
\end{equation*}
observations. The indicator function $\one_A(x)$  returns the value $1$ if $A$ holds true for $x$ and the value $0$ otherwise. A judicious choice of $h$ ensures a large enough number $N(\theta_j)$  is present, so that the extreme value theorem holds on $\mathcal{N}(\theta_j, h)$, for every $\theta_j \in \Theta$. The largest $k_j$ observations in $\mathcal{N}(\theta_j)$ are then taken as threshold exceedances for the direction-specific estimation of $\xi(\theta_j)$ and the optimal number of threshold exceedances is the number $k^*_j$ satisfying $S_\phi(k^*_j)=\minover{k} S_\phi(k_j)$ with:
\begin{equation}\label{autoChoice_k}
S_\phi (k)= \frac{1}{k}\sum_{i \leq k} i^\phi \bigl| \hat{\xi}_{i}(\theta_j) - \textrm{median}\bigl(\hat{\xi}_{1}(\theta_j), \hat{\xi}_{2}(\theta_j), \dots, \hat{\xi}_{k}(\theta_j)\bigr) \bigr|,
\end{equation}
where $0 \leq \phi < 0.5$, and $\hat{\xi}_{k}(\theta_j)$ stands for the designated estimator of $\xi(\theta_j)$ restricted to the $k$ upper observations amongst the $N(\theta_j)$ neighboring observations defined by $\mathcal{N}(\theta_j, h)$. The heuristic procedure \eqref{autoChoice_k} was introduced by Reiss and Thomas \citep[cf.]{reiss2007statistical} and then studied in detail in \citet{NFA2004}. It facilitates an automatic choice of the threshold which can be understood intuitively as follows. For small $k$, the weighted deviations in \eqref{autoChoice_k} tend to be large due to the inherently large variance of $\hat{\xi}_k(\theta_j)$. As $k$ increases, the summands in \eqref{autoChoice_k} are expected to decrease until bias sets in and overrides the variance from which point $S_\phi$ is expected to increase again. Minimizing the weighted empirical distance \eqref{autoChoice_k} is equivalent to optimizing the bias-variance trade-off by exploiting the settled behaviour of estimates $\{\hat{\xi}_k:\, k < N\}$ for appropriate $k$. A plethora of estimators for $\xi$ has been proposed in the literature, most notably, the moment estimator and parametric maximum likelihood estimator. These two estimators are extended here to directional estimators that not only rely on the magnitude of excesses above a threshold but also, and perhaps more critically, take into account the directional spread of exceedances relative to their centroid $\theta_j$.

The motivation for the automatic selection of a moving threshold across $\theta_j \in \Theta$ is to couple threshold exceedances originating at every location $\theta_{j'} \in \mathcal{N}(\theta_j,h) \cap \Theta$ with their propensity $\omega(\theta_{j'})$ for spreading around $\theta_j \in \Theta$. Since we are dealing with circular data, the von Mises kernel is a natural choice for measuring this spread \citep[cf.][]{PewseyNR14}. Precisely, we define weights:
\begin{equation}\label{NWweight}
\omega(\theta_{j'}) := \frac{ \mathcal{K}_\eta(\theta_{j'})}{\sumab{\theta_{j'} \in \mathcal{N}(\theta_j, h)}{} \mathcal{K}_\eta( \theta_{j'})}, 
\end{equation}
with the von Mises kernel implicitly defined on the centroid $\theta_j \in \Theta$ as:
\begin{equation*}
\mathcal{K}_\eta(s) := \frac{1}{2 \pi B_0(\eta)}\,\exp \big\{\eta \cos (s- \theta_j) \big\},
\end{equation*}
for $s\in \mathcal{S}$. The modified Bessel function of the first kind of order zero, $B_0(\eta)= \pi^{-1}\int_0^\pi e^{\eta \cos s}\, ds$, gives the required normalization in order to  ensure $\mathcal{K}_\eta$ is in fact a density function. The concentration parameter $\eta > 0$ in the von Mises kernel controls the spread of the kernel in the sense that the greater the value of $\eta$, the greater the concentration and the lower the spread of the exceedances about the centroid $\theta_j\in  \Theta$. Hence, this parameter plays a role similar to bandwidth $h$ intervening in \ref{NeigbDirect}, with both quantities playing out as important contributors to the degree of smoothness in the adaptive threshold estimation through the Moment (M) and maximum likelihood  (ML) estimators given below in Sections~\ref{Sct:ThrEst:SP} and \ref{Sct:ThrEst:ML}.

\subsection{Local semi-parametric estimation} \label{Sct:ThrEst:SP}
For each $\theta_j \in \Theta$, the extended version of the Moment estimator for $\xi(\theta_j)$ embedding directional weights \eqref{NWweight} is given by
\begin{equation}\label{MomMovThrKer}
\hat{\xi}^{M}_k(\theta_j) \coloneqq M^{(1)}(\theta_j) + 1 - \frac{1}{2}\Big(1 - \frac{\bigl(M^{(1)}(\theta_j) \bigr)^2}{M^{(2)}(\theta_j)}\Big)^{-1},
\end{equation}
with 
\begin{equation*}
M^{(l)}(\theta_j) \coloneqq \sum_{ \substack{\theta_{j'}\in \mathcal{N}(\theta_j,h)\\ i =1, \ldots, n}  } \omega(\theta_{j'}) \big(\log X_{i}(\theta_{j'})- \log X_{N(\theta_j)\,-k, N(\theta_j)} \big)^l\,\one_{\{X_{i}(\theta_{j'}) > X_{N(\theta_j)\,-k, N(\theta_j)} \}} , \qquad l=1,2,
\end{equation*}
where $X_{N(\theta_j)\,-k, N(\theta_j)} $ denotes the $(k+1)^\text{th}$ largest value in the observed sample, whose directional covariate $\theta_{j'}\in \Theta$ belongs to $\mathcal{N}(\theta_j,h)$.

This framework is key to the semi-parametric approach. The operative assumption relates to the asymptotic behavior of the $k$-th upper order statistics associated with the sample of i.i.d. positive random variables $X_i(\theta_j')-X_{N(\theta_j)\,-k, N(\theta_j)}$ established in the theory of extremes for threshold excesses. Conditionally on $X_{N(\theta_j)\,-k, N(\theta_j)}=u$, the common distribution function for these random variables is $F^{[u]}(t)= P\bigl(X(\theta_j)-u >t |\,X(\theta_j)>t \bigr)$, for $t>0$ \citep[cf.][page 90]{deHaan2006extreme}. See  Appendix \ref{Sct:Append} for a precise definition of $F^{[u]}$, and how it approaches the GPD function.

\subsection{Local maximum likelihood estimation} \label{Sct:ThrEst:ML}
Along similar lines, the local directionally-weighted ML estimator $\hat{\xi}_k(\theta_j)$, for every $\theta_j \in \Theta$, is the result of maximizing, with respect to the parameter-vector $ \bigl( \xi(\theta_j), \sigma_u(\theta_j)\bigr)\in (-1, \infty) \times \real^+$, the weighted log-likelihood:
\begin{equation}\label{KernelLogLikNonStat}
L\bigl(\xi(\theta_j), \sigma_u(\theta_j)\bigr)  \coloneqq   \sum_{ \substack{\theta_{j'}\in \mathcal{N}(\theta_j,h)\\ i =1, \ldots, n}  } \omega(\theta_{j'}) \, \ell \big(\xi(\theta_j), \sigma_u(\theta_j) |\,X_i(\theta_{j'})-X_{N(\theta_j)\,-k, N(\theta_j)} \big) \one_{\{X_i(\theta_{j'})-X_{N(\theta_j)\,-k, N(\theta_j)} >0 \}},
\end{equation}
with weights $\omega(\theta_{j'})$ as in \eqref{NWweight}, and:
\begin{equation*}
\ell \big(\xi(\theta_j), \sigma_u(\theta_j)|\,y\big) =  - \log \sigma_u(\theta_j) - \big(1+1/\xi(\theta_j)\big) \log \big( 1+\xi(\theta_j)\, y/ \sigma_u(\theta_j)\big)
\end{equation*}
when $\xi(\theta_j) \neq 0$. For $\xi(\theta_j) = 0$, a continuity argument yields
\begin{equation*}
\ell \big(\sigma_u(\theta_j) | \,y \big) = -\log \sigma_u(\theta_j) -y/\sigma(\theta_j).
\end{equation*}
This ML formulation has been tailored for heuristic threshold selection, which itself has the semi-parametric method at its core. The link with parametric ML estimation is seen since, conditioned on the random number of exceedances $K=k$, random exceedances of threshold $X_{N(\theta_j)\,-k, N(\theta_j)}$ can be viewed as i.i.d. random variables with distribution function $F^{[u]}$ \citep[details are deferred to part (i) of Appendix~\ref{Sct:Append}; cf. e.g.][page 234]{reiss2007statistical}.
\subsection{Algorithm for non-stationary threshold selection} \label{Sct:ThrEst:Alg}
For each $\theta_j$, the local estimates $\hat{\xi}_k(\theta_j)$, $k=1, \ldots, N(\theta_j)-1$ used in \eqref{autoChoice_k} and resulting $k_j^*$, will reflect the extent of stationarity on $\mathcal{N}(\theta_j,h)$.  The procedure described above for optimal selection of non-stationary threshold for all $\theta_j \in \Theta$ is outlined in Algorithm \ref{NPalg}. We do not claim that this heuristic approach to extreme value threshold specification is optimal for every estimator one might devise for the extreme value index $\xi$, but we have found it useful in the applications context of Section~\ref{Sec:App}.

\begin{algorithm}[h]
	\caption{Non-stationary threshold estimation}
	\begin{algorithmic}[1]
		\State Specify lag $h$, concentration $\eta$ and parameter $\phi$;
		\For{$\theta_j$ $\in \Theta$}
		\State Estimate weights $\omega(\theta_{j'})$ for $\theta_{j'}$ $\in \Theta$;
			\For{$k=1,\dots,N(\theta_j)-1$}
			\State Use the $k$ largest values in  $\mathcal{N}(\theta_j,h)$ to estimate $\hat{\xi}_{k}(\theta_j)$ using \eqref{KernelLogLikNonStat} or \eqref{MomMovThrKer};
			\State Calculate $S_\phi(k)$;
			\EndFor
		\State Set $k_j^*=\argmin{k}\,S_\phi(k)$;
		\State Identify threshold $u(\theta_j)$ with the $(k_j^*+1)^{\text{th}}$ largest value with direction in $\mathcal{N}(\theta_j,h)$;
		\EndFor

	\end{algorithmic}
	\label{NPalg}
\end{algorithm}


\section{Spline-based maximum likelihood estimation}\label{Sct:SplML}
Given a non-stationary threshold $u(\theta_j)$, $\theta_j \in \Theta$, such as that obtained in Section~\ref{Sct:ThrEst}, we proceed with parametric peaks over threshold analysis. We take the sample of threshold exceedances identified above, using the local maximum likelihood approach in Section~\ref{Sct:ThrEst:ML}, and perform further parametric extreme value analysis. The purpose of this extra inference step is to mimic a conventional analysis of non-stationary threshold exceedances that might be undertaken in ocean engineering (see e.g. \cite{northrop2016thre}). Specifically, we assume a B-spline representation for the variation of GPD shape and scale parameters with covariate. We estimate spline coefficients using maximum penalized likelihood estimation, regulating the roughness of shape and scale with covariate to optimize predictive performance assessed using cross-validation. In Section~\ref{Sct:ExtQnt} we use the fitted model to infer $T$-year levels and (if appropriate) right endpoint.

In the interest of physically meaningful inference, we assume that the shape and scale parameters vary smoothly with respect to covariate $\theta \in \mathcal{S}$, adopting smooth functions for $\xi(\theta)$ and $\log \sigma(\theta)$ using periodic cubic B-spline basis functions on $\mathcal{S}$ (see e.g. Chapter 5 of \citealt{wood2017generalized}, \citealt{ZnnEA19a}). On the index set $\Theta$ of covariate values $\theta_j, \ j=1,2,\dots,m$, we relate the values $\xi(\theta_j), \sigma(\theta_j)$ of GPD shape and scale to the periodic B-spline basis via basis matrix $\un{B}$ with elements $B_{jb}$ such that:
\begin{equation}\label{bsplines}
\xi(\theta_j) = \sum_{b=1}^{n_b} B_{jb} \beta_b^{(1)} \quad \text{ and } \quad
\log \sigma(\theta_j) = \sum_{b=1}^{n_b} B_{jb} \beta_b^{(2)} , \qquad \theta_j \in \Theta,
\end{equation}
where $n_b$ is the number of basis functions, and the $\beta$s are basis coefficients. The sample log-likelihood is:
\begin{equation}\label{LogLikNonStat}
\ell(\un{\beta}) = - \sum_{j=1}^m \sum_{i=1}^{n_j} \Big( \log \sigma(\theta_j) + \big(1+\frac{1}{\xi(\theta_j)}\big)  \log \big(1 + \frac{\xi(\theta_j)}{\sigma(\theta_j)} (X_i(\theta_j)-\widehat{u}(\theta_j) \ \big)\Big) \one_{\{X_{i}(\theta_j) > \widehat{u}(\theta_j)\}},
\end{equation}
with $n_j$ denoting the number of observations in the sample at covariate $\theta_j$, $\un{\beta}^{(a)}=(\beta_1^{(a)},\beta_2^{(a)},\dots,\beta_{n_b}^{(a)})^\top$ for $a=1,2$, and $\un{\beta}=(\un{\beta}^{(1)\top},\un{\beta}^{(2)\top})^\top$. We set the number of spline knots on $\mathcal{S}$ to be more than sufficient to capture the anticipated parameter variability with covariate, and then penalize parameter roughness globally to obtain a model with good predictive performance. penalization in performed using first-order difference penalties for the coefficients in \eqref{bsplines},
\begin{equation*}
P^{(a)} = \sum_{b=1}^{n_b-1} \Big(\beta_{b+1}^{(a)} - \beta_{b}^{(a)}\Big)^2 + \Big(\beta_{1}^{(a)} - \beta_{n}^{(a)}\Big)^2 = \boldsymbol{\beta}^{(a)\top} \boldsymbol{D}^\top \boldsymbol{D} \boldsymbol{\beta}^{(a)}, \qquad a =1,2,
\end{equation*}
with difference matrix $\un{D}$ given by:
\begin{equation*}
\boldsymbol{D} = \begin{bmatrix} 
-1 & 1 & 0 & 0 & \cdot & \cdot \\
0 & -1 & 1 & 0 & \cdot & \cdot \\
\cdot & \cdot & \cdot & \cdot & \cdot & \cdot \\
\cdot & \cdot & \cdot & \cdot & \cdot & \cdot \\
0 & 0 & \cdot & 0 & -1 & 1 \\
1 & 0 & \cdot & 0 & 0 & -1
\end{bmatrix}.
\end{equation*}
The penalized log-likelihood is then:
\begin{equation}\label{PenLogLikNonStat}
\ell_{pen}(\boldsymbol{\beta}) = \ell(\boldsymbol{\beta}) - \lambda P^{(1)} - \kappa P^{(2)},
\end{equation}
where $\lambda$ and $\kappa$ are smoothing parameters chosen maximise cross-validated predictive likelihood. In the application illustrated in Section \ref{Sec:App}, cross-validation is applied as follows. Each iteration of the cross-validation consists in using a bootstrap resample (of the original sample of threshold exceedances) as training set, and observations omitted from the bootstrap resample as test set. Note that the test sets corresponding to different bootstrap resamples may therefore overlap. The training set is used to estimate the model parameters, and the test set to assess prediction performance using mean squared prediction error (MSPE). The procedure is outlined in Algorithm~\ref{Palg}.

\begin{algorithm}
	\caption{Maximum penalized likelihood for non-stationary periodic B-spline representation}
	\begin{algorithmic}[1]
		\State Evaluate B-spline basis functions on index set $\Theta$;
		\State Specify sets of values of smoothing penalty coefficients $\lambda$ and $\kappa$ to consider;
		\For{each choice of $\lambda$}
		\For{each choice of $\kappa$}
		\For{each of a number of bootstrap resamples}
		\State Estimate model parameters using bootstrap resample;
		\State Use estimated model to predict test observations (not occurring in bootstrap resample);
		\State Calculate the mean squared prediction error;
		\EndFor
		\State Accumulate (average) mean squared prediction error;
		\EndFor
		\EndFor
		\State Select pair of values of $\lambda$ and $\kappa$ with best predictive performance, and evaluate spline coefficients for these choices of roughness coefficients;	
	\end{algorithmic}
	\label{Palg}
\end{algorithm}

\section{ Estimation of a non-stationary extreme quantile}\label{Sct:ExtQnt}
This section is devoted to the estimation of an extreme quantile, including that of the right endpoint in the case of a short-tailed distribution ($\xi<0$). We first present a class of maximum likelihood estimators for high quantiles in the parametric setting, and then introduce adapted versions of widely-used semi-parametric estimators amenable to the moving threshold $u(\theta)$, $\theta \in \mathcal{S}$, introduced in Section~\ref{Sct:ThrEst}. An algorithmic guide to inference of extreme quantiles and finite right endpoint, using any of the estimators discussed here, is given in Appendix~B.

In this section we assume a known (fixed, given) deterministic threshold $u(\theta)$, $\theta \in \mathcal{S}$ is available, as estimated in Section \ref{Sct:ThrEst:Alg}. Exceedances are taken above threshold $X_{N(\theta)\,-k(\theta), N(\theta)}=u(\theta)$, where $k(\theta)$ is the number of exceedances in neighborhood $\mathcal{N}(\theta, h) \subset \mathcal{S}$ defined in \eqref{NeigbDirect}. Further, let $k(\theta)/N(\theta)$ be the top sample fraction which will be retained for inference on extreme values, and $N(\theta)$ the total the number of observations in neighborhood $\mathcal{N}(\theta, h)$ for direction $\theta$.
Given $u(\theta)$, the resulting random exceedances are distributed as $K=k(\theta)$ i.i.d. random variables with the same distribution function $F^{[u(\theta)]}$. This setting entails that any information in $X_{N(\theta)\,-k(\theta), N(\theta)}$ is disregarded \citep[cf.][page 90]{deHaan2006extreme}, and fits ideally to the parametric POT-GPD framework.
	
Let $F_\theta$ be the actual distribution function underlying data measurements taken at direction $\theta \in \mathcal{S}$ and let $Q_\theta$ be the corresponding quantile function, i.e. $Q_\theta=F_\theta^\leftarrow$ with the left arrow indicating a generalized inverse. The basic theory for extremes (see Appendix \ref{Sct:Append}) establishes the GPD function as the limit for the distribution of the linearly normalized exceedances over a threshold. In fact, conditions \eqref{POTparam} and \eqref{POTapprox} ensure that an extreme quantile $F^\leftarrow(1-p)$, with $1-p > F(u)$ for some high threshold $u$,  only depends on the tail of the distribution function $F$. Consequently, it will be possible to estimate an extreme quantile associated with a very small probability $p$ via the corresponding linear functional $\hat{b} + \hat{a} \,Q^H(1-p)$, with $Q^H$ now pertaining to the GP distribution function $H_{\xi}(x)= 1-(1+\xi x)^{-1/\xi}$, for all $x>0$ and $1+\xi x >0$, $\xi \in \real$. This setting carries over to both parametric and semi-parametric approaches, although with slight and yet potentially impactful differences. The most obvious of these is that following a parametric approach, the normalising constants $a$ and $b$ are ascribed to scale and location of the limiting GPD, whereas in the semi-parametric approach these are estimated as functionals of the sample analogue to the distribution function $F_{\theta}$. 

In the parametric setting, the level $x_p$ with small exceedance probability $p$ corresponds to the quantile of the distribution of the POT-exceedances at direction $\theta$. This suggests defining the $1/p$ extreme quantile as the value $x_{p}= Q^H(1-p)$, with normalizing constants $a>0$ and $b$ set to scale and location parameters of the GPD function $H$ for large $1/p$. Hence the local ML and spline ML estimator of $x_{p}$:
\begin{equation}\label{RLparametric}
\widehat{x}_{p}^{ML}(\theta) = u(\theta) + \hat{\sigma}(\theta)\, \frac{\bigl(\frac{\varphi(u,\theta)}{p} \bigr)^{\hat\xi(\theta)}-1}{\hat\xi(\theta)},
\end{equation}
with $\varphi(u,\theta)=k(\theta)/N(\theta)$, the sample fraction of exceedances of $u(\theta)$ within $\mathcal{N}(\theta,h)$. The estimator of an extreme level in \eqref{RLparametric} can be interpreted as a $T$-year level used in hydrology and ocean engineering: $x(T)$ is that value exceeded on average once a year, i.e. $P\bigl(X> x(T)\bigr)= 1/T$, whereby we put $x(T)=x_{1/T}= Q^H(1-1/T)$.

Local ML estimates for parameters $\xi(\theta_j)$ and $\sigma(\theta_j)$ on the index set $\theta_j \in \Theta$ for \eqref{RLparametric} are obtained using the likelihood criterion in Section \ref{Sct:ThrEst}, now with constant weights $\omega(\theta_{j'})$ in \eqref{KernelLogLikNonStat}. The values of $k(\theta_j)$ and $N(\theta_j)$ are estimated in Section~\ref{Sct:ThrEst}.

In the spline ML approach of Section \ref{Sct:SplML}, $\varphi(u,\theta_j)$ is estimated as the probability of threshold exceedance for $\theta_j \in \Theta$ using logistic regression, with log-likelihood:
\begin{equation}\label{EllBeta}
	\ell(\un{\beta}) = \sum_{j=1}^m \tau_j \log[\nu_j] + (1-\tau_j) \log[1-\nu_j],
\end{equation}
where $\tau_j$ is the sample proportion of threshold exceedances of $u$ at $\theta_j$. $\nu_j=(1+\exp[-\eta_j])$ is the probability of threshold exceedance at $\theta_j$, with $\un{\eta} = \{\eta_j\}_{j=1}^m = \un{B}\un{\beta}$ for B-spline basis matrix $\un{B}$ and parameter vector $\un{\beta}$ to be estimated. Roughness penalization of $\un{\eta}$, with optimal roughness coefficient $\mu$ estimated by cross-validation, ensures good predictive performance. The penalized log-likelihood thus takes the form $\ell_{pen}(\un{\beta}) = \ell(\un{\beta}) - \mu \un{P}$, with $\un{P}=\un{\beta}^\top \un{D}^\top \un{D} \un{\beta}$ (see Section~\ref{Sct:SplML}). Estimation of $\xi(\theta)$ and $\sigma(\theta)$ for the spline ML approach is explained in Section~\ref{Sct:SplML}.

In case $\xi(\theta)<0$, the limiting GP distribution function  has a finite right endpoint which we also seek to estimate. A consistent estimator for this right endpoint follows readily from \eqref{RLparametric} by setting $p=0$:
\begin{equation}\label{EPparametric}
\hat{x}_{0}(\theta)= {u}(\theta) - \frac{\hat{\sigma}(\theta)}{\hat{ \xi}(\theta)}.
\end{equation}

In the semi-parametric setting, we assume that the intermediate number $k(\theta)$ of exceedances above threshold $u(\theta)$ is such that $k(\theta) \rightarrow \infty$ and $k(\theta)/N(\theta) \rightarrow 0$ as $N(\theta)\rightarrow \infty$. That is, the number of threshold exceedances $k(\theta)$ remains negligible compared to the total number $N(\theta)$ of observations in neighborhood $\mathcal{N}(\theta,h)$. The latter can be rephrased in terms of the direction-specific sample size $n$ for the theoretical underpinning to domains of attraction that requires  $n\rightarrow \infty$ also implies $N(\theta) \rightarrow \infty$. The proposed M estimator for the extreme quantile with small probability $p=p_n \rightarrow 0$ and $N(\theta)p/k(\theta) \rightarrow 0$, as $n \rightarrow \infty$, conditioned on the threshold $u(\theta)$ now follows:
\begin{equation}\label{HQsemiparametric}
\widehat{x}^{M}_{p}(\theta) = 
u(\theta) + \widehat{a}_{\theta}^{M}\bigl( \frac{N(\theta)}{k(\theta)}\bigr)\, \frac{\Big( \frac{k(\theta) }{N(\theta) p}\Big)^{\hat{\xi}^{M}_{k}(\theta)}-1}{\hat{\xi}^{M}_{k}(\theta)},
\end{equation}
with $k(\theta)$ assumed fixed. We then proceed in a semi-parametric way with the plug-in M estimator for the EVI defined in the same way as before in \eqref{MomMovThrKer}:
\begin{equation}\label{MomentEstPar}
	\hat{\xi}^{M}_k(\theta_j) \coloneqq M^{(1)}(\theta_j) + 1 - \frac{1}{2}\Big(1 - \frac{\bigl(M^{(1)}(\theta_j) \bigr)^2}{M^{(2)}(\theta_j)}\Big)^{-1},
\end{equation}
and associated scale estimator:
\begin{equation}\label{afctEst}
\hat{a}^M_{\theta}\Bigl(\frac{n}{k(\theta)}\Bigr):= u(\theta)\, M^{(1)}(\theta) \frac{1}{2} \Big(1 - \frac{(M^{(1)}(\theta))^2}{M^{(2)}(\theta)}\Big)^{-1},
\end{equation}
with $M^{(l)}(\theta)$, $l=1,2$, updated to:
\begin{equation*}
M^{(l)}(\theta) = \frac{1}{k(\theta)}\sumab{ i=1}{k(\theta)-1}\sumab{\theta_{j} \in \mathcal{N}(\theta,h)}{}  \big(\log X_{i}(\theta_j)- \log {u}(\theta_j)\big)^l\,\one_{\{X_{i}(\theta_j) > {u}(\theta_j)\}}.
\end{equation*}
Expressions \eqref{RLparametric} and \eqref{HQsemiparametric} show obvious similarities, and also distinctive traits of the M, local ML and spline ML approaches. Scale $a>0$ in \eqref{HQsemiparametric} is a function of the top sample fraction $k(\theta)/N(\theta)$, compared with the sample-free parameter GPD $\sigma>0$ in \eqref{RLparametric}. In view of theoretical development in Appendix \ref{Sct:Append}(ii), the estimator of the scale function $\hat{a}^M$ refers to a consistent estimator $\hat{a}_{\theta}$ for the scale function $a^*_\theta\bigl(\nicefrac{n}{k \omega(\theta)} \bigr)$ therein.

The class of estimators \eqref{EPparametric} for the right endpoint $x^F$ has been much studied from parametric and semi-parametric perspectives \citep[cf. Chapter 4 of][and connected references]{deHaan2006extreme}. We now propose a model-free, data-driven estimator of $x^{F}(\theta)<\infty$, $\theta \in \mathcal{S}$, motivated by the general endpoint estimator of  \citet{FAN2014}, coupled with non-stationary threshold $u(\theta)$. Specifically, we formulate an extreme value condition aiming to induce a partition of the Gumbel max-domain of attraction into a class of distributions with finite upper bound and a class containing the remainder. An example of a distribution function in the former class is the Negative Fr\'echet, with distribution function $F_{\alpha,\beta}(x)= 1-\exp\{-(\alpha-x)^{-\beta}\}$, $x \leq \alpha$, $\alpha \in \real$, $\beta>0$. Simple calculations show that $F_{\alpha,\beta}$ belongs to the Gumbel max-domain of attraction (hence with $\xi=0$) despite having finite right endpoint $x^F=\alpha<\infty$. In this sense, a semi-parametric estimator is likely to fare better than the parametric alternative in connection with small negative values of the shape parameter $\xi$. The parametric approach will tend to bear on the archetypical exponential distribution for inference (i.e., drawing on the POT-GDP with $\xi=0$ and infinite upper bound) and hence distributions associated with $\xi=0$ albeit with finite right endpoint will escape its grasp.

With $\bigl\{Y_{i}\bigr\}_{i=1}^{N(\theta)}$ representing the random variables $X_i$s in a particular neighborhood $\mathcal{N}(\theta,h)$, defined in \eqref{NeigbDirect}, we denote by $Y_{1,N(\theta)}\leq  \ldots \leq Y_{N(\theta)-k(\theta),N(\theta)}\leq \ldots\leq Y_{N(\theta),N(\theta)}$ the corresponding ascending  order statistics, and define the general endpoint estimator of $x^{F(\theta)}$, assumed finite, for every $\theta \in \mathcal{S}$,
\begin{equation}\label{generalEndpoint}
\hat{x}_0^G(\theta) :=  Y_{N(\theta),N(\theta)} + \frac{1}{\log 2}\sumab{i=0}{k(\theta)-1} \log \Bigl(\frac{k(\theta)+i+1}{k(\theta)+i} \Bigr)  \big( Y_{N(\theta)-k(\theta),N(\theta)} - Y_{N(\theta)-k(\theta)-i,N(\theta)}\big).
\end{equation}
Notably, this estimator is valid for any $\xi(\theta)\leq 0$ deeming any prior estimation of the EVI unnecessary \citep[cf.][]{FANR17}.

\section{Application to the storm peak significant wave height}\label{Sec:App}
The methodology above provides an approach to applied non-stationary extreme value analysis incorporating elements from both parametric and semi-parametric inference as described. A key feature of the methodology is the estimation of a non-stationary threshold capturing the covariate dependence of large values of response so as to balance the number and spread  of large observations on the covariate domain as explained in Section~\ref{Sct:ThrEst}. In this section, we apply the methodology to estimation of $T$-year  values from the sample of storm peak significant wave height $H_S^{sp}$ on storm direction $\theta$ introduced in Section~\ref{Sct:Dat}. The mechanics of inference for extreme quantiles and right endpoint, if appropriate, is given in Appendix~\ref{Sct:AppendOutline}.

\begin{figure}
	\centering\texttt{}
	\includegraphics[scale=0.25]{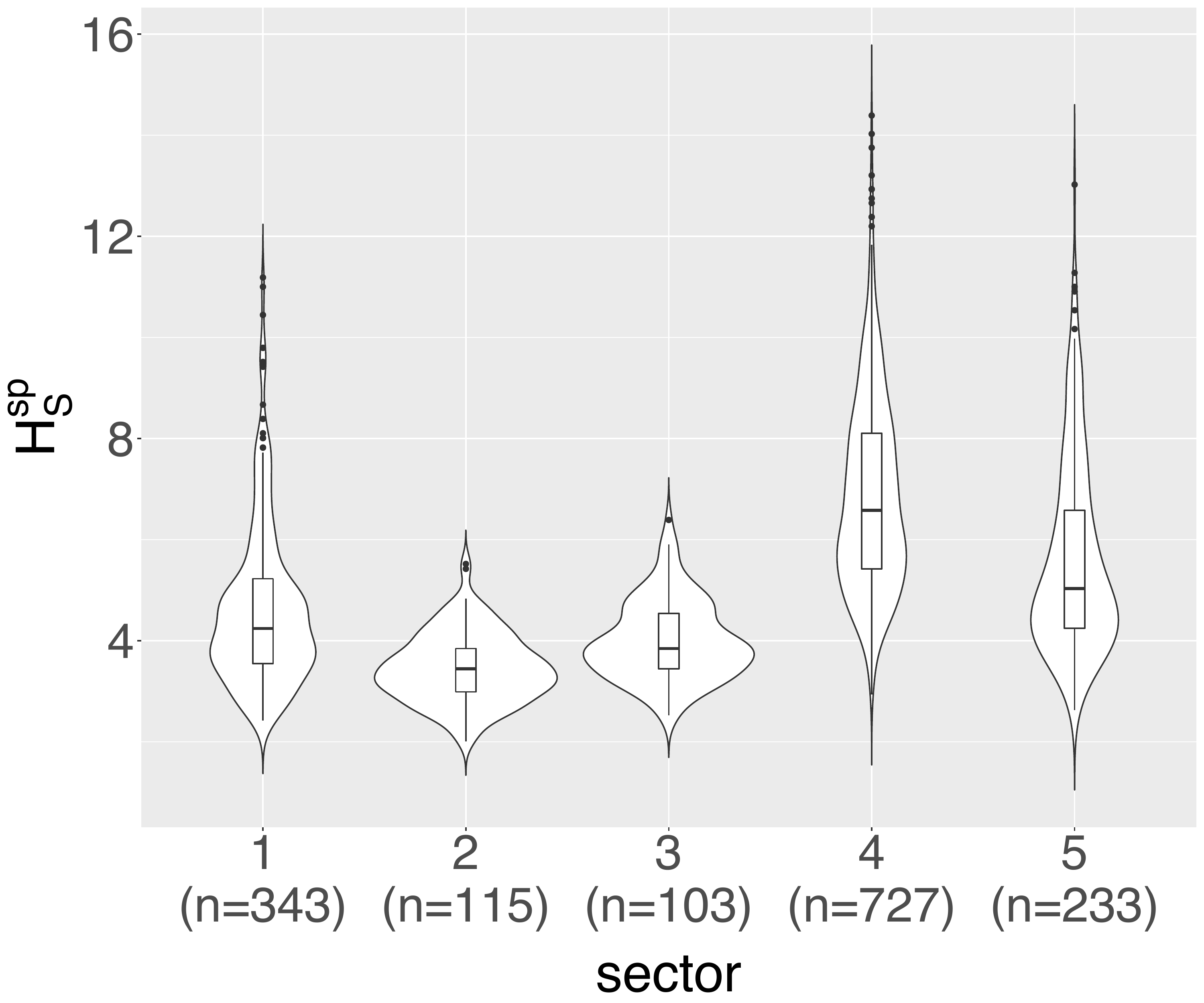}
	\caption{~Violin plots of \HsSP{} observations to the total of $N=1521$ data points.}
	\label{Fig:ViolinData}
\end{figure}

Exploratory analysis of the sample, supported by previous analysis by \cite{RndEA15a}, suggests that the covariate domain can be partitioned into five directional sectors assumed approximately homogeneous in terms of the characteristics of $H_S^{sp}$. Referring to Figure~\ref{Fig:HsSPdata}, directional sectors corresponding to the following intervals of $\theta$ were identified. Sector 1 corresponds to $\theta \in (0^\circ,50^\circ] \cup (321^\circ,360^\circ]$, for storms propagating from the Norwegian Sea to the North; Sector 2 for $\theta \in (50^\circ,140^\circ]$ corresponds to the ``land shadow'' of Norway, with fetch-limited storms propagating from the coast with a more northerly direction relative to the normal to the coast; Sector 3 is $\theta \in (140^\circ,210^\circ]$, again for the Norwegian land-shadow, but with storms propagating from a more southerly direction; Sector 4 is $\theta \in (210^\circ,270^\circ]$ corresponding to storms from the Atlantic potentially ``funnelled'' by the Norwegian coast; and Sector 5 with $\theta \in (270^\circ,320^\circ]$, for more northerly Atlantic storms. Further information about the underlying physics is given in Section~\ref{Sct:Dat}. The partitioned sample is summarised in Figure~\ref{Fig:ViolinData}, using so-called ``violin'' plots which add kernel density estimates to a box-whisker representation. The long-tailed behaviour of storms from the Atlantic is clear in Sectors 4 and 5, compared to the fetch-limited characteristics in storms from Sectors 2 and 3. Although Sector 4 exhibits the largest values of threshold exceedances in Figure~\ref{Fig:ViolinData}, there is evidence from the kernel density plots that Sector 5 has a relatively long tail. In this section we seek to quantify tail-heaviness by estimating EVI / shape parameter $\xi$ using both parametric and semi-parametric approaches, and hence estimate extreme quantiles.

\begin{figure}
	\centering
	\includegraphics[scale=0.5]{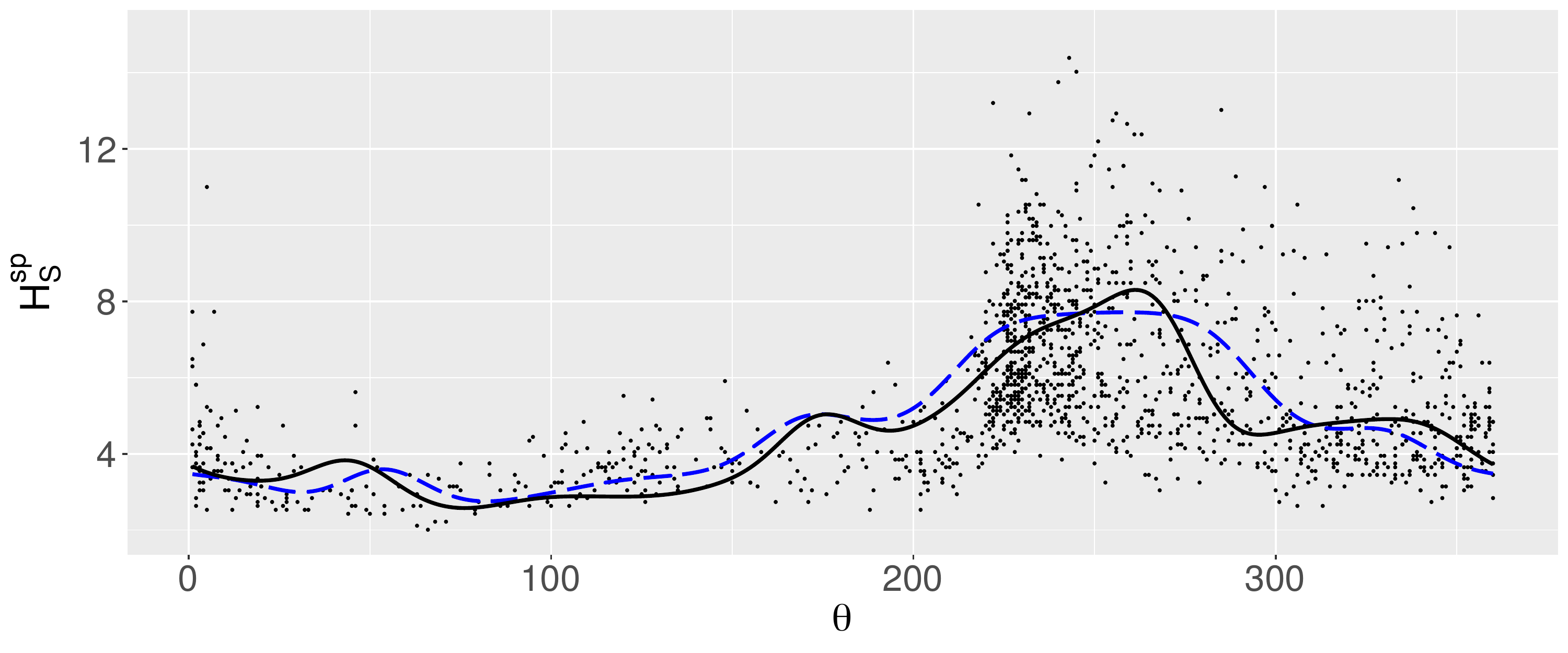}
	\caption{~Adaptive threshold selection on the basis of the parametric ML estimator (black, solid line) and the semi-parametric M estimator with lag $h=30^\circ$, concentration $\eta=10$ and parameter $\phi=0.35$.}
	\label{Fig:thresholds}
\end{figure}

Using the approach in Section~\ref{Sct:ThrEst}, non-stationary thresholds were estimated using the (parametric) ML estimator, and the (semi-parametric) M estimator for lag $h=30^\circ$, concentration $\eta=10$ and parameter $\phi=0.35$. Estimates are shown together in Figure~\ref{Fig:thresholds}. The general trends shown by the two estimates are in good agreement across the covariate domain. Subsequent parametric and semi-parametric inference for exceedance characteristics therefore has a relatively common starting point.

\begin{figure}
	\centering
	\includegraphics[scale=0.18]{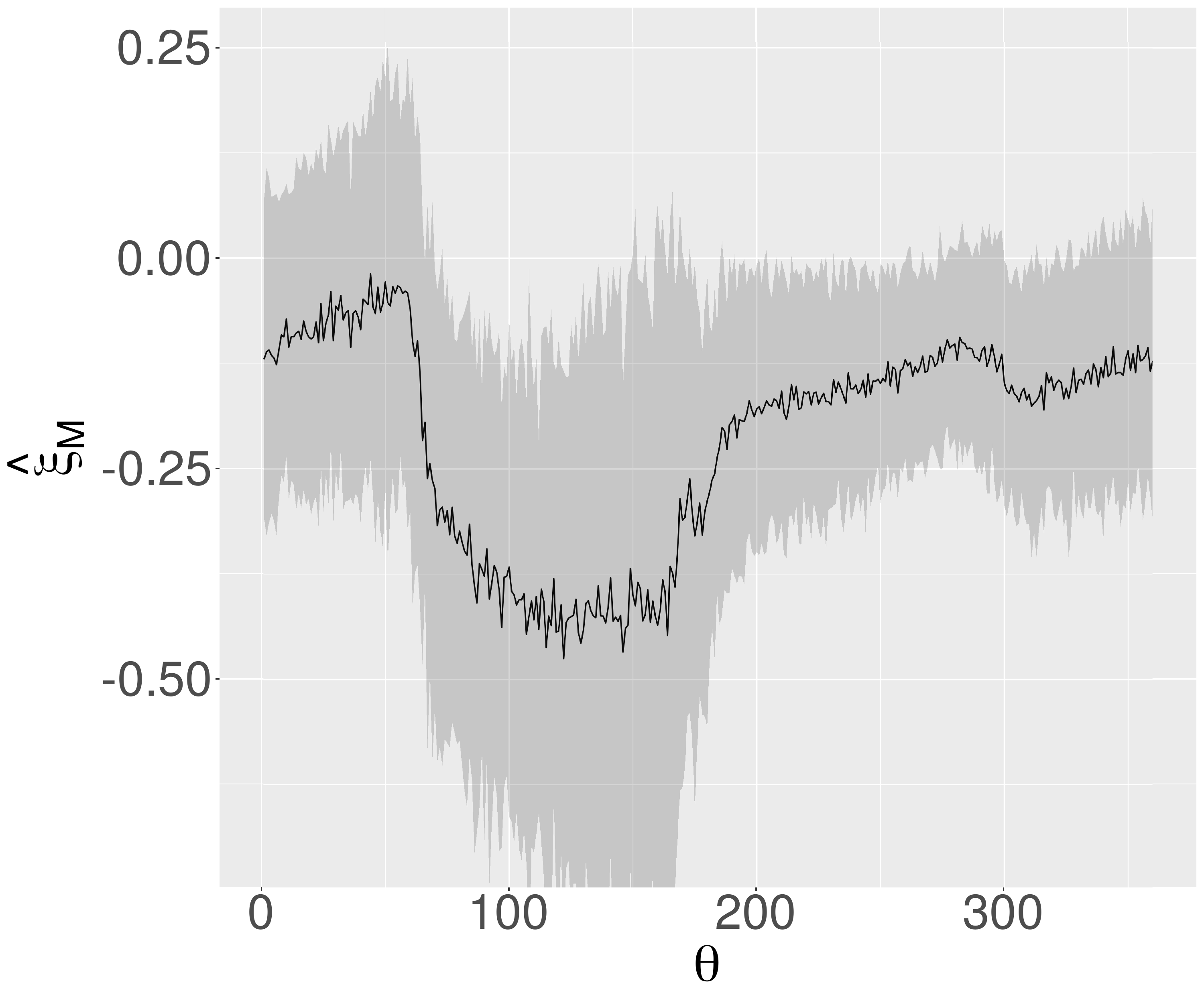}
	\includegraphics[scale=0.18]{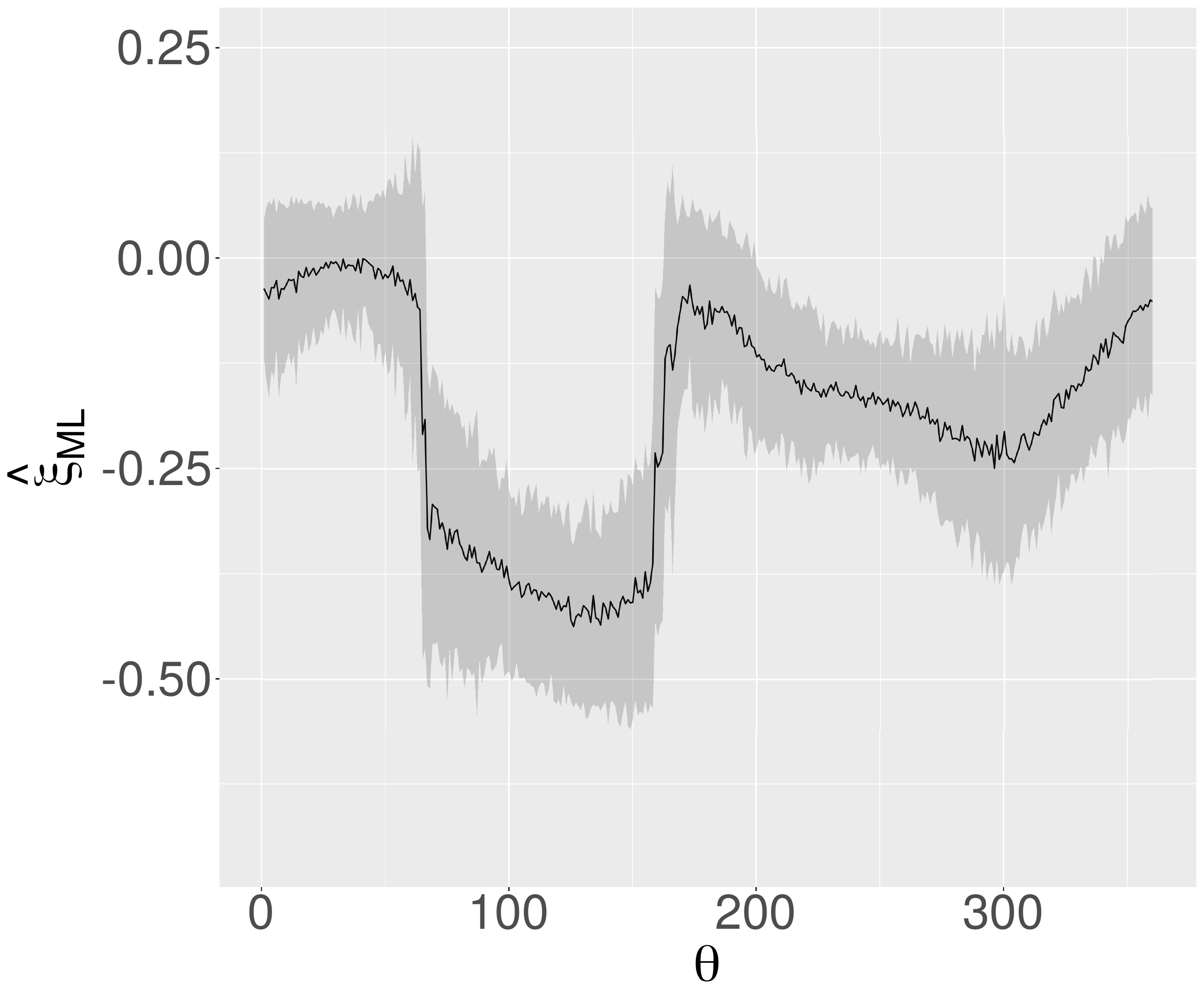}
	\includegraphics[scale=0.18]{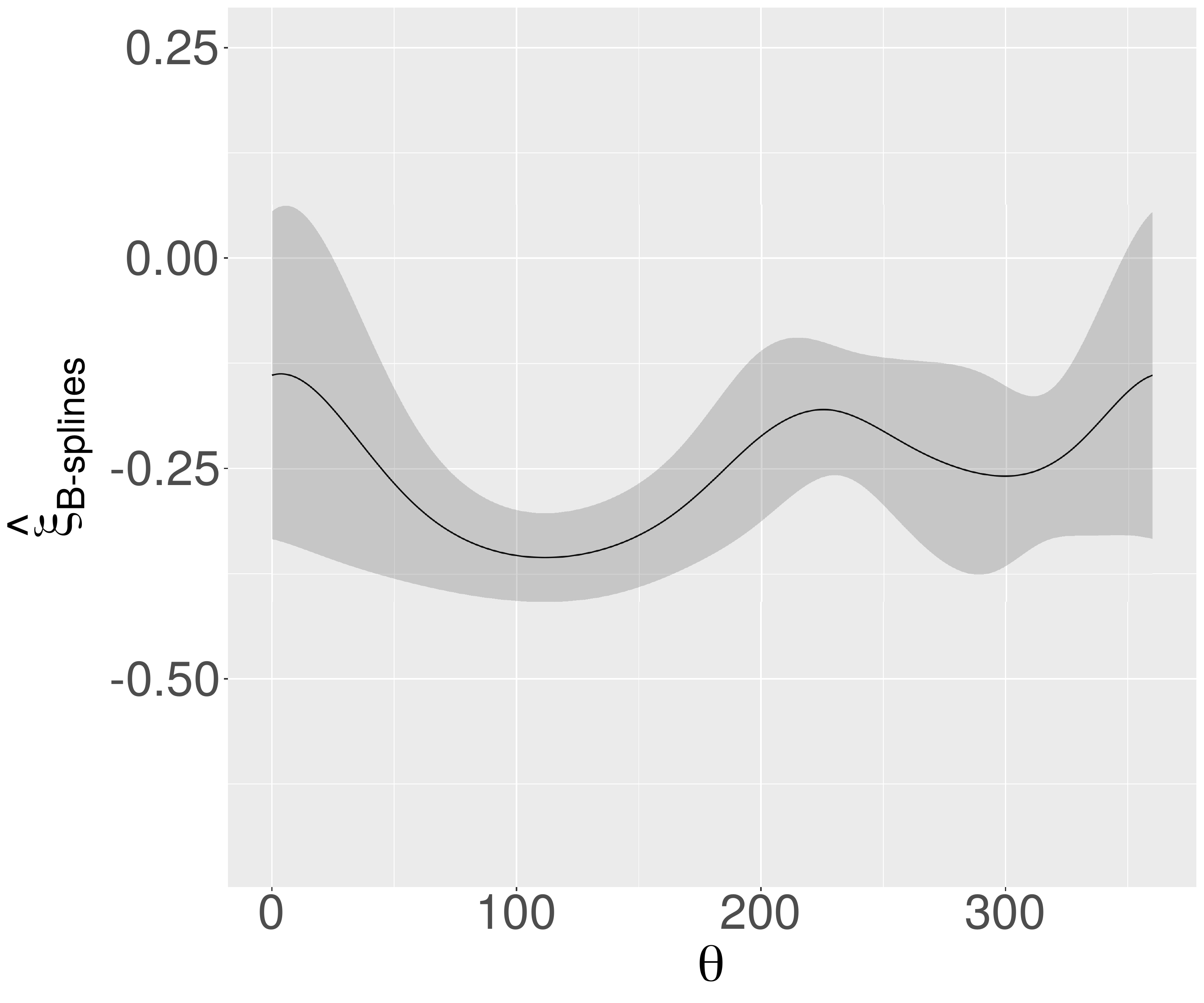}
	\caption{~Bootstrap mean and corresponding 95\% confidence intervals for $\xi(\theta_j)$, $\theta_j \in \Theta$, based on: M (\emph{left}), local ML (\emph{middle}) and spline ML (\emph{right}) estimators.}
	\label{Fig:EVI}
\end{figure}

Estimates $\hat{\xi}(\theta_j)$ of EVI or GPD shape parameter $\xi$, gauging tail heaviness, for every centroid $\theta_j \in \Theta$ from each of the M, local ML an spline ML approaches is displayed in Figure~\ref{Fig:EVI}, in terms of bootstrap means and 95\% confidence intervals. Overall, there is good qualitative agreement between the three estimates. The estimates are also qualitatively plausible given other analyses of these data (\citealt{RndEA15a}) and physical considerations. Effects of land shadows (e.g. $\theta \in (80,150)$) resulting in low $\xi$ are clear. The M estimator exhibits wider confident intervals. The spline ML estimator is smoothest with respect to covariate. Local ML and spline ML estimates make an additional asymptotic GPD assumption which the M estimate does not. The local ML estimator is in some senses intermediate, and this is reflected in the figure. Confidence limits exceed zero for all estimates, but this is more pronounced with the M estimation. Indeed, both M and local ML estimates for EVI do not differ significantly from zero for a large part of the covariate domain, suggesting that the data generating distribution lies in the Gumbel max-domain of attraction there.

\begin{figure}
	\centering
	\includegraphics[scale=0.32]{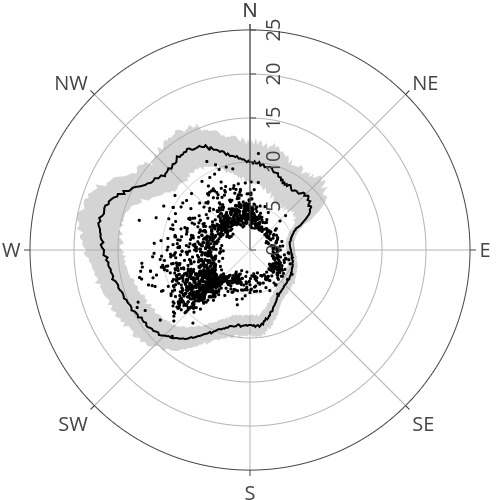}
	\includegraphics[scale=0.32]{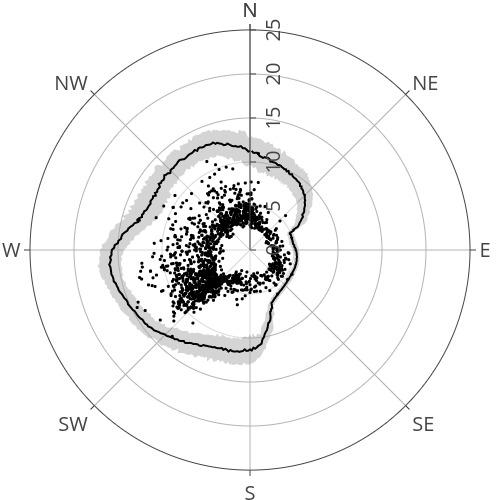}
	\includegraphics[scale=0.32]{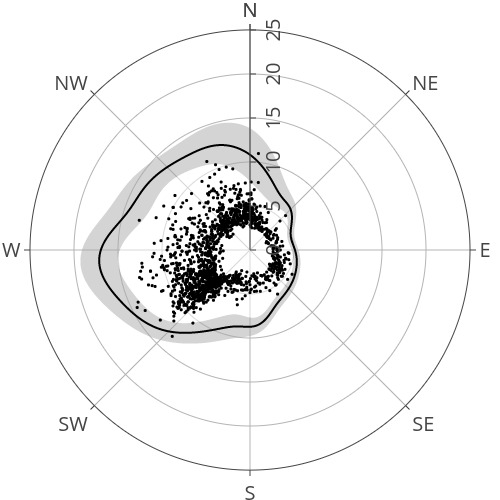}
	\includegraphics[scale=0.32]{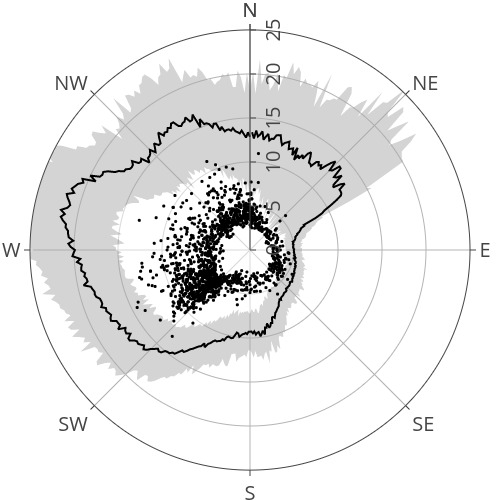}
	\includegraphics[scale=0.32]{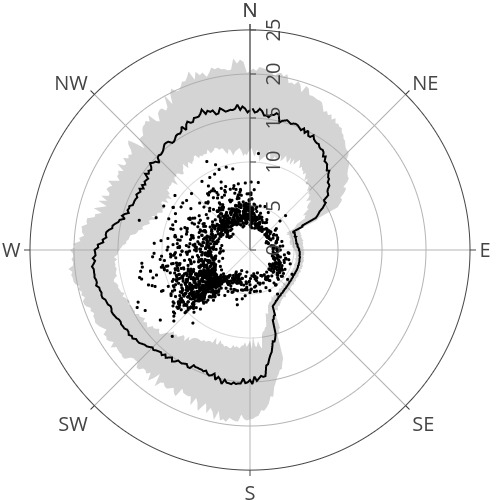}
	\includegraphics[scale=0.32]{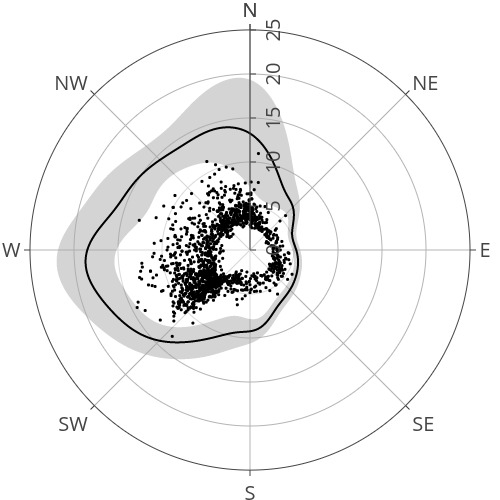}
	\caption{~Rose diagrams for the $100$-year (top) and $10,000$-year (bottom) levels based on M (\emph{left}), local ML (\emph{middle}) and spline ML (\emph{right}) estimators. Bootstrap means and corresponding 95\% confidence bands are displayed as a function of direction measured clockwise from north.}
	\label{Fig:Tyearevent}
\end{figure}

Estimation of $T$-year levels is straightforward once relevant POT-GPD parameters are estimated. Using expressions \eqref{RLparametric} and \eqref{HQsemiparametric}, assuming $N_E$ occurrences of storm in observation period $T_0$, the $T$-year directional level corresponds to the value $x_p(\theta)$ such that $P\{ X(\theta)  > x_p(\theta) \} = (T_0/N_E) \times T^{-1}$. Figure~\ref{Fig:Tyearevent} displays in a matrix rose-plots for the estimated $100$-year and $10,000$-year levels with accompanying 95\% bootstrap confidence bands. Again, there is general qualitative agreement between the three estimates for $100$-year level  (top row) and $10,000$-year level (bottom row), in terms of bootstrap mean. Uncertainties from the M estimate are somewhat larger, as might be expected recalling the evidence of Figure~\ref{Fig:EVI}. Not surprisingly, estimated extreme levels for directions with short fetches ($\theta \in [70,140)^\circ$) are low, whereas those corresponding to long fetches from the Atlantic Ocean and Norwegian Sea ($\theta \in [225,360)\cup0,40)^\circ$) are high. Estimated $100$-year return levels for \HsSP{} fall between 15m and 20m the most severe sectors, in terms of bootstrap mean and confidence bands. The same is true of bootstrap means at the $10,000$-year level except for M estimates which exceed 20m. This is not inconsistent with evidence from Figure~\ref{Fig:EVI} regarding generally negative $\xi$ estimates.

\begin{figure}
	\centering
	\includegraphics[scale=0.32]{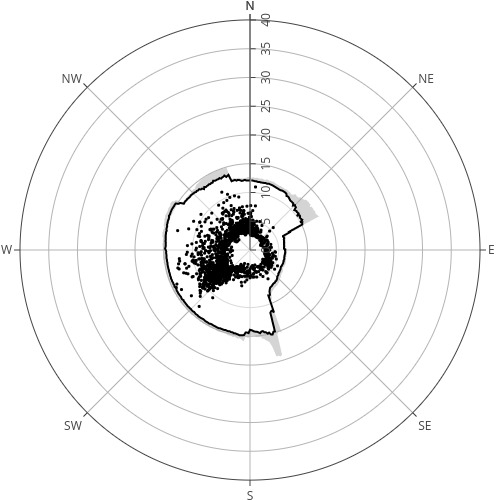}
	\includegraphics[scale=0.32]{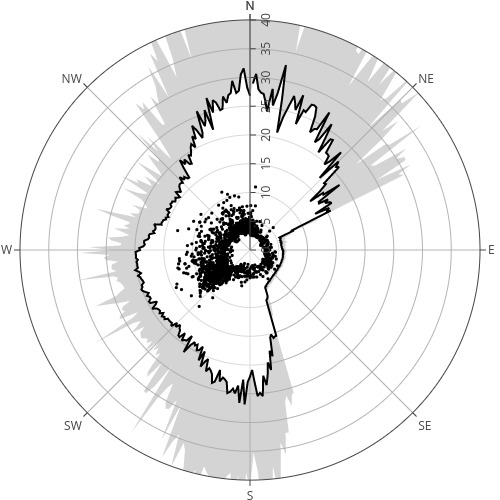}
	\includegraphics[scale=0.32]{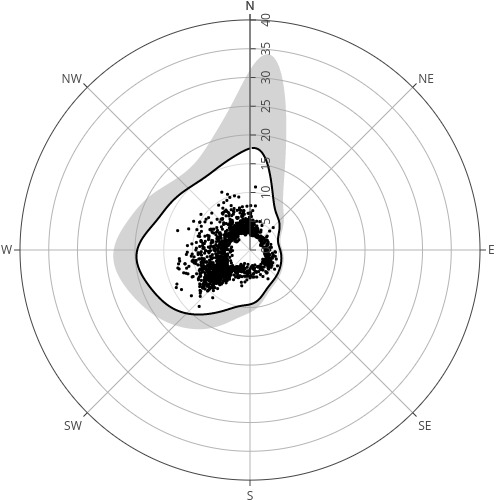}
	\caption{~Rose diagrams for the finite right endpoint of \HsSP{}. Three estimators are used: the general estimator (\emph{left}), and the local ML (\emph{middle}) and spline ML (\emph{right}) estimators. Bootstrap means and corresponding 95\% confidence bands are displayed as a function of direction measured clockwise from north.}
	\label{Fig:EP}
\end{figure}

\begin{figure}
	\centering
	\includegraphics[scale=0.35]{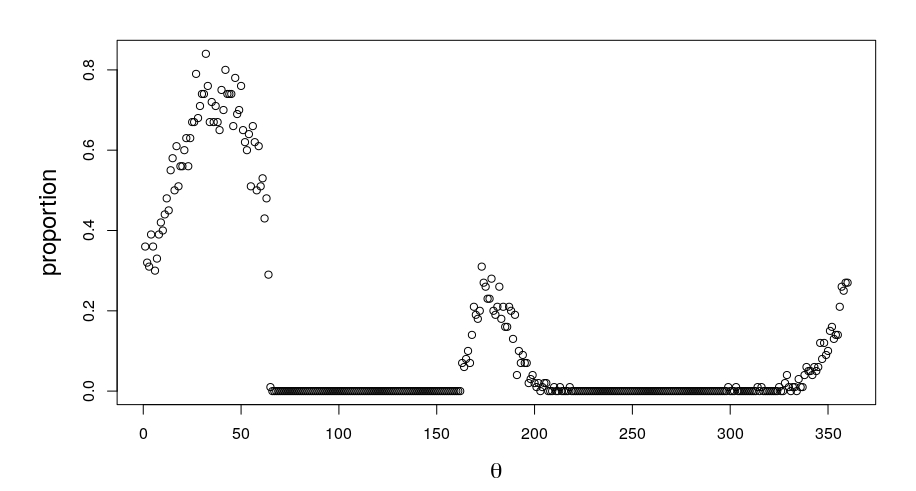}
	\includegraphics[scale=0.35]{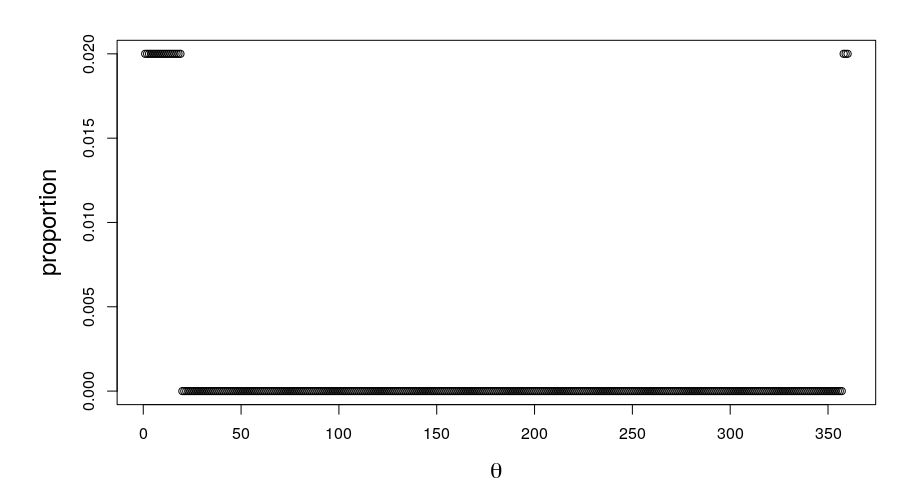}
	\caption{~Fraction of bootstrap samples excluded from the finite right endpoint estimation in Figure \ref{Fig:EP} for failing to return a negative estimate of $\xi$, for (left) local ML and (right) spline ML.}
	\label{Fig:Prop}
\end{figure}

Assuming that threshold exceedances are indeed drawn from a short-tailed distribution independent of direction, we can also consider estimation of the largest possible value of $H_S^{sp}$ given the sample. Inspection of Figure~\ref{Fig:EVI} suggests that $\xi$ exceeds zero for all estimators for some values of covariate, and therefore it may be that the right endpoint is infinite there. Figure~\ref{Fig:EP} shows estimates corresponding to the general estimator \eqref{generalEndpoint} for the finite right endpoint on the covariate domain, which does not require any estimation of $\xi$.  Note that the general endpoint estimation does not rest on any value of the shape parameter/EVI, but rather on the less stringent assumption is that the distribution underlying the data has a finite endpoint. Figure~\ref{Fig:EP} also shows estimates of the right endpoint using outputs of local ML and spline ML estimation in \eqref{EPparametric}, applied only when the estimated value of $\xi$ is negative. Figure~\ref{Fig:Prop} shows the proportion of bootstrap samples excluded from the ML-based inference, since $\hat\xi\geq 0$. ML estimation becomes more challenging as the true shape parameter $\xi$ approaches zero from negative values, with numerical optimization routines more than often experiencing convergence issues \citep[see e.g.][for a comparison between M and ML within the univariate semi-parametric setting]{GomesNeves2008}. 

Agreement between the three estimators for the finite right endpoint is not as good as that observed for $T$-year level. The characteristics of estimates from the general estimator and spline ML are in some agreement, apart from for northerly directions, for which spline ML suggests that a longer tail is present (and see Figure~\ref{Fig:Prop} for $\hat\xi>0$ from the North). For the Atlantic sector, local ML estimates are also in relatively good agreement with the others; however, local ML provides large numbers of estimates for $\xi$ exceeding zero for northerly and southerly storms. This produces large estimates for the right endpoint with large uncertainties. Indeed, were we to have attempted to estimate the right endpoint using semi-parametric M estimates for $\xi$ directly in \eqref{EPparametric}, resulting endpoint estimates would also have been large and highly variable since a large proportion of M estimates for $\xi$ are fairly close to zero (cf. Figure~\ref{Fig:EVI}).

In summary, application of the parametric and semi-parametric methodologies developed in Sections~\ref{Sct:ThrEst}-\ref{Sct:ExtQnt} above to the sample of directional storm peak significant wave height suggests that estimates for extreme value index/ GPD shape, and $100$-year and $10,000$-year levels are in good qualitative agreement. Where differences occur, they can be understood and explained in terms of specific modelling assumptions made, rather than in terms of fundamental differences in the underlying approaches to extreme value analysis.

\section{Summarizing remarks}\label{Sec:Summary}

This paper presents a framework for inference on non-stationary peaks over threshold, reconciling approaches from semi-parametric and parametric extreme value analysis, in application to directional ocean storm severity. The key components of the framework are (a) estimation of non-stationary extreme value threshold, and (b) estimation of tail characteristics from threshold exceedances, including extreme quantiles and right endpoint when appropriate (finite). Threshold estimation is performed using a non-stationary extension of a heuristic approach proposed by \citet{NFA2004} for semi-parametric moment (M) and parametric maximum likelihood (ML) estimators.

Tail characteristics and extreme quantiles are then estimated, based on semi-parametric M, local parametric ML and spline ML estimators. We also develop a non-stationary semi-parametric general endpoint estimator (based on \citet{FAN2014} for $\xi \leq 0$), and apply it with the standard right endpoint estimator (applicable for $\xi<0$) where appropriate.

Inferences regarding directional thresholds for storm peak significant wave height are in good agreement over the covariate domain. Estimates for $100$ and $10,000$-year levels are also in reasonable agreement. Estimates for the right endpoint are more different across approaches, and are influenced by the specifics of modelling assumptions made associated with the different estimation strategies. For the application considered, both parametric and semi-parametric inference provides similar characterisations of extreme non-stationary ocean environments. Indeed, we illustrate how ideas from the semi-parametric and parametric schools of thought can be used in tandem to exploit the desirable features of the approaches, whilst overcoming some obvious pitfalls. For example, threshold estimation (used for both semi-parametric and parametric analysis) is motivated by an inherently non-parametric heuristic in Section~\ref{Sct:ThrEst}. 

Parametric approaches to non-stationary extremes are relatively well-studied due in part to the wide range of flexible covariate representations for GPD parameters for threshold exceedances, and associated methods for regression and assessment of model fit. In contrast, from a semi-parametric perspective, a tangible GPD exact fit need not be assumed, but avoiding a particular model choice from the outset generally results in increased uncertainty of estimates for EVI and high quantiles. However, we show in this work that semi-parametric and parametric approaches perform rather similarly when set up reasonably.

By exploiting recent developments of extreme value theory for non-identically distributed observations \citep[cf.][]{deHaan2015tail,deHaanZhou20}, we show that under reasonable and mildly restrictive assumptions, a suitable number of parameters can be introduced with the aim of optimizing the for bias/variance trade-off in the estimation of the various extreme value characteristics and/or indices. This optimization method relies heavily on an adaptive choice for the non-stationary threshold which determines where tail-related observations begin to show up in the available sample. Given threshold, the two streams of development mirror each other regarding tail inference. In the spline ML approach, a cubic B-splines representation with compact support is used: each basis function is non-zero on a specific interval of the covariate domain. This feature plays a similar role to bandwidth of directional neighborhoods in the semi-parametric M and local ML approaches. We anticipate that the framework presented here can be extended to address multidimensional covariates often encountered in practice. 

\subsection*{ACKNOWLEDGEMENT}
Cl\'audia Neves gratefully acknowledges support from EPSRC-UKRI Innovation Fellowship grant EP/S001263/1 and project FCT-UIDB/00006/2020.

\subsection*{ORCID}
\emph{Evandro Konzen}  \orcidA{} \url{https://orcid.org/0000-0002-6275-1681}\\
\emph{Cl\'audia Neves} \orcidB{} \url{https://orcid.org/0000-0003-1201-5720}\\
\emph{Philip Jonathan} \orcidC{} \url{https://orcid.org/0000-0001-7651-9181}

\bibliography{NonStatRefereces2020}
\bibliographystyle{apalike}

\appendix

\section{Basis for inference on directional extremes via the POT method}\label{Sct:Append}

The contents of this appendix build on Appendix B of \citet{deHaan2015tail}. This paper adds flexibility to the latter because we give allowances for $\xi=\xi(s)$ to vary with $s \geq 0$, where $s$ might represent direction $\theta$, or time or some other covariate. As a consequence, the right endpoint does not need to be assumed constant in $s$. We will not delve into the theoretical details in terms of explicit smoothness and boundedness conditions needing to be in place particularly by assuming $h=h_n>0$. These are clearly beyond the scope of this paper, but we envisage that the probabilistic underpinning to this work will stem from Chapter 9 of \citet{deHaan2006extreme}.

At direction (or time) $s \geq 0$, let $\bigl( X_1(s), X_2(s), \ldots, X_n(s)\bigr)$ be a vector of i.i.d. random variables with common distribution function $F_s(x)$, for all $x \in \mathbb{R}$, absolutely continuous with right endpoint $x^{F_s} \leq \infty$.  Assume $F_s \in \mathcal{D}(G_{\xi(s)})$, for some $\xi(s) \in \mathbb{R}$ and for every $s\geq 0$, i.e., that condition \eqref{DOA} holds locally for each $s$. In this setting, Theorem 1.1.6 of \citet{deHaan2006extreme} ascertains that it is possible to replace $n$ with $t$ running over the real line in such a way that \eqref{DOA} becomes equivalent to the following extreme value condition: there exists a positive function $a^*_s$ such that
\begin{equation}\label{POTdomain}
	\lim_{t \uparrow x^{F_s}}\, \frac{ 1-F_s\bigl( t + x a^*_s(t)  \bigr)}{1-F_s(t)} = \frac{\log G_{\xi(s)}(x)}{\log G_{\xi(s)}(0)},
\end{equation}
for all $x$ with $1+ \xi(s) x >0$. The limit in \eqref{POTdomain} is the tail distribution function (also known as survival function) of the GP distribution with shape parameter $\xi(s) \in \real$, given by $\bigl(1+ \xi(s) x\bigr)^{-1/\xi(s)}$. The extreme value condition \eqref{POTdomain} is often key for describing rare events' behavior in lieu of the dual max-domain of attraction characterization $F\in \mathcal{D}(G_\xi)$.

\subsection*{(i) Parametric approach}
Taking a parametric view, the POT-domain of attraction condition \eqref{POTdomain} prescribes the GP distribution as the proper fit to the normalized exceedances given these are above a certain high threshold near the right endpoint $x^{F_s}$.
With minimal notational changes around a fixed (deterministic) threshold $u$, it is straightforward to see that condition \eqref{POTdomain} implies 
\begin{equation}\label{POTparam}
	\lim_{u \uparrow x^{F_s}} \, \Bigl|P\{X_1(s) \leq x+u \, |\, X_1(s) > u\} - H_{\xi(s), u, \sigma(u)}(x) \Bigr|=0,
\end{equation}
locally uniformly in $x> u$, for each $s\geq 0$, with $\sigma_s(u)>0$ (henceforth we omit the subscript $s$ for simplicity of notation) and $	H_{\xi, \mu, \sigma}(x):= 1-	 \bigl(1+\xi (x-\mu)/\sigma\bigr)^{-1/\xi}$,
for all $x$ such that $0<H_{\xi, \mu, \sigma}(x)<1$, with location $\mu \in \mathbb{R}$ and scale $\sigma>0$.  Informally, $F_s^{[u]}(x):= P\{X_1(s) \leq x+u \, |\, X_1(s) > u\} \approx H_{\xi(s), u, \sigma(u)}(x)$, for all $x> 0$ and $u$ near the right endpoint $x^{F_s}$, with the scale parameter implicitly defined in terms of $s$ through the threshold $u(s)$. 

For each $s$, the limiting relation \eqref{POTparam} provides the probabilistic underpinning for fitting a GP distribution function to the unconditional tail distribution function $\overline{F_s}(x):= 1-F_s(x)$ with $x$ sufficiently large. This becomes more evident since
\begin{eqnarray*}
	& & F_s(x) = (1- F_s(u)) F_s^{[u]}(x) + F_s(u),
	\end{eqnarray*}
whence,
\begin{equation}\label{POTapprox}
	\overline{F_s}(x)= \bigl(1-F_s^{[u]}(x)\bigr)(1- F_s(u)) \approx \bigl(1-   H_{\xi(s), u, \sigma(u)}(x)\bigr)(1- F_s(u)),
\end{equation}
for $ x > u$, as $u \rightarrow x^{F_s}$.  Finally, we note that
\begin{equation}\label{Rep}
	\bigl(1-   H_{\xi(s), u, \sigma(u)}(x)\bigr)(1- F_s(u))= H_{\xi(s), \mu^*, \sigma^*(u)}(x),
\end{equation}
where $\mu^*-u= \sigma(u) \,U_{H} \bigl(1- F_s(u)\bigr)$, $\sigma^*(u) = \sigma(u) \bigl(1- F_s(u)\bigr)^{\xi(s)}$, and $U_H$ standing for the tail quantile function pertaining to the standard GPD, that is
\begin{equation*}
	U_H(t):= \biggl(\frac{1}{1-H_{\xi(s), 0, 1}}\biggr)^{\leftarrow}(t)=\frac{t^{\xi(s)}-1}{\xi(s)},
\end{equation*}
for all $t \geq 1$ (the left arrow indicates the left-continuous inverse). The  representation \eqref{Rep} facilitates the view that, in practice, changes in the threshold (e.g. through covariates) will be reflected in the scale parameter. In turn, the approach for inference is reflected in the way we go choose to go about the term $1- F_s(u)$ for this to become statistically meaningful. In order to able to perform large sample inference drawing on the POT-GPD framework streamlined above, we now make the threshold dependent on the sample size $n$ and $u=u(n)$ will naturally become larger as $n$ goes to infinity. A parametric approach typically advocates for a large enough threshold to be fixed and inference to be conducted on the basis of the resulting POT framework, whereby the expected number of exceedances above the selected threshold is a random number $K_s$, say, satisfying $(n/K_s)(1- F_s(u_n)) \rightarrow 1$ in probability, as $n\rightarrow \infty$. This suggests estimation of $(1- F_s(u))$ via the analogous tail empirical distribution function (stepping up by $1/n$ at each observation) evaluated at $u$, adding up to $1-K_s/n$ in the above, associated with the random number $K_s$ of exceedances of $u$ at direction (or location) $s$. Hence, for a given (fixed) $k_s$, the location and scale parameters in \eqref{Rep} become:
\begin{eqnarray*}
	\mu^*(s)&=& u + \sigma(u) \,U_{H} \Bigl(1- \frac{k_s}{n}\Bigr),\\
	\sigma^*(u) &=& \sigma(u) \Bigl(1- \frac{k_s}{n}\Bigr)^{\xi(s)}.
\end{eqnarray*}
Therefore, the crux of parametric inference for extremes values lays in the estimation of the shape and scale parameters, respectively, $\xi(s)$ and $\sigma_s=\sigma^*(u)$.

\subsection*{(ii) Semi-parametric approach}

It will be notationally cleaner to express the argument \eqref{POTdomain} in terms of the pertaining tail quantile function $U_s:= \bigl(1/(1-F_s)\bigr)^\leftarrow$. Note that $U_s(t)$ is non-decreasing and provides a straightforward link to an extreme quantile, with the right endpoint representing the ultimate quantile: $\lim_{t\rightarrow \infty} U_s(t)=U_s(\infty)= x^{F_s}$. To this effect, we make $t$ in \eqref{POTdomain} depend on the (possibly unknown) sample size $n$ at each location $s$ through replacing it by $U_s(n/k_s)$, where $k_s$ is an intermediate sequence of positive integers such that $k_s=k_s(n) \rightarrow \infty$ and $k_s/n \rightarrow 0$, as $n\rightarrow \infty$. This is possible because \eqref{POTdomain} holds uniformly in $x$. Hence, we have for the left hand-side of \eqref{POTdomain}:
\begin{equation*}
	\frac{1-F_s\bigl(U_s(n/k_s)+ x \, a_s\bigl(U_s(n/k_s)\bigr) \bigr)}{1-F_s\bigl(U_s(n/k_s)\bigr)} =  \frac{n}{k_s}\bigl(1-F_s\bigl(U_s(n/k_s)+ x \, a^\star_s(n/k_s) \bigr),
\end{equation*}
with $a_s^\star(n/k_s)=a_s\bigl(U_s(n/k_s)\bigr)$.

For simplicity, we consider regularly spaced independent vectors $\bigl(X_{1}(s), X_{2}(s), \ldots, X_{n}(s)\bigr)$, $s=1,2, \ldots, m, \ldots$ with i.i.d. components, with partial tally of $N=n\times m \in \mathbb{N}$ observations across the whole system, and where $n$ is potentially unknown (without affecting inference on extremes), yet assumed large ($n\rightarrow \infty$). In this setting, the basic extreme value condition is:
\begin{equation}\label{DOAnonstat}
	\lim_{n \rightarrow \infty} \frac{N}{\omega(s) k}\Bigl[1-F_s\Bigl(\,U_s\bigl(\frac{n}{\omega(s)  k}\bigr)+ x \, a^{\star}_s\bigl(\frac{n}{\omega(s)  k}\bigr) \Bigr) \Bigr]= \bigl(1+\xi(s)x\bigr)^{-1/\xi(s)},
\end{equation}
for all $x$ with $1+\xi(s)x>0$, uniformly in $s=1, 2, \ldots$, subject to $(1/m)\sum_{s= 1}^m \omega(s) \rightarrow 1$, as $m\rightarrow \infty$, i.e. the sequence of weights $\{\omega(m)\}_{m\in \field{N}}$ is Ces\`{a}ro summable. The latter is to maintain integrity, also ensuring that the stationary case is well-defined. In particular, the case of complete stationarity, corresponding to omni-directional data in the context of this paper, is recovered if $\omega(s)$ is uniformly distributed over the stipulated range for $s$. The interest lies in the estimation of the various extreme value indices $\xi(s)$, and the scale and location terms, respectively $a_s^\star(n/k_s)$ and $U_s(n/k_s)$, now with $k_s:= [\omega(s) \times k]$ and $[\bigcdot]$ standing for integer part. Since the $n$-th order statistic $X_{n-k_s:n}(s)$ is close to $U_s(n/k_s)$, if $k_s \rightarrow \infty$, $k_s/n \rightarrow 0$, as $n \rightarrow \infty$, we shall adopt it, as the usual estimator for the threshold $\widehat{U}_s(n/k_s)=X_{n-k_s:n}(s)$. The random adaptive threshold in this setting emulates the non-stationarity mirrored in the scale $\sigma_s>0$ which features the parametric setting (i). 

\section{.~Roadmap for application}\label{Sct:AppendOutline}

The procedure for estimation of an extreme level and of the right endpoint, if appropriate, for any of the three methods is summarised in the following algorithm for clarity. The algorithm assumes that the non-stationary threshold $u(\theta)$, $\theta \in \mathcal{S}$ is already known.

\begin{algorithm}
	\caption{Extreme quantile (including $T$-year level) and finite right endpoint estimation using M, local ML and spline ML based estimators.}
	\begin{algorithmic}[1]
		\State Specify data sample; non-stationary threshold $u(\theta)$; period of sample $T_0$, return period $T$;
		\State Isolate set of directional threshold exceedances;
		\If{M}
			\State Specify window half-width $h$;
		\ElsIf{local ML}
			\State Specify window half-width $h$;		
		\ElsIf{spline ML}
			\State Specify values of smoothing coefficients $\lambda$ and $\kappa$ to consider;
			\State Specify details for B-spline basis function construction;
		\EndIf
		\For{each of a large number of bootstrap resamples}
	
			\State Generate bootstrap resample from sample of threshold exceedances;
			\If{M}
				\State Count number $k(\theta)$ of threshold exceedances in $\mathcal{N}(\theta,h)$;
				\State Count number $N(\theta)$ of observations in $\mathcal{N}(\theta,h)$ ;
				\State Estimate $\hat\xi(\theta)$ on $\mathcal{N}(\theta,h)$ using \eqref{MomentEstPar};
				\State Estimate $\hat{a}_\theta$ on $\mathcal{N}(\theta,h)$ using \eqref{afctEst};
				\State Estimate high quantile with $p << 1/N(\theta)$ value using \eqref{HQsemiparametric};
				\State Estimate finite right endpoint ($x^{F_\theta}<\infty$) using \eqref{generalEndpoint};
			\ElsIf{local ML}
				\State Count number $k(\theta)$ of threshold exceedances in $\mathcal{N}(\theta,h)$;
				\State Count number $N(\theta)$ of observations in $\mathcal{N}(\theta,h)$ ;
				\State Estimate $\hat\xi(\theta)$, $\hat\sigma(\theta)$ on $\mathcal{N}(\theta,h)$ (using (4.6) with weights $\omega=1$);
				\State Estimate $T$-year return value using \eqref{RLparametric};
				\State Estimate right endpoint (when $\hat\xi(\theta)<0$) using \eqref{EPparametric};
			\ElsIf{spline ML}
				\State Estimate optimal smoothing parameters $\lambda$, $\kappa$ and hence estimate $\hat\xi(\theta)$, $\hat\sigma(\theta)$ (Algorithm 2);
				\State Estimate optimal smoothing parameter $\mu$ and fraction $\hat\tau(\theta)$ of threshold exceedances by logistic regression using \eqref{EllBeta};
				\State Estimate $T$-year return value using \eqref{RLparametric}, with $\hat\tau(\theta)$ in place of $k(\theta)/N(\theta)$;
				\State Estimate right endpoint (when $\hat\xi(\theta)<0$) using \eqref{EPparametric};		
			\EndIf
			
			\State Accumulate bootstrap estimates for parameters, extreme levels or quantiles and for the right endpoint;
				
		\EndFor
		\State Calculate bootstrap means and confidence intervals for parameters, return values and endpoint;
\end{algorithmic}
\end{algorithm}

\end{document}